\documentclass[10pt]{article}
\usepackage[a4paper,
            margin=1in]{geometry}
\usepackage{parskip}
\usepackage[activate={true,nocompatibility},final,kerning=true,spacing=true,factor=1100,stretch=10,shrink=10]{microtype}
\usepackage{setspace}
\usepackage{amsmath}
\usepackage{amssymb}
\usepackage{graphicx}
\usepackage{booktabs}
\usepackage{sectsty}
\usepackage{tabularx}
\usepackage[normalem]{ulem}
\usepackage{csquotes}
\usepackage[htt]{hyphenat}
\usepackage{nameref}
\usepackage{mathabx}
\usepackage{totcount}
\usepackage{epigraph}
\usepackage[hyphens]{url}
\usepackage{hyperref}
\usepackage{todonotes}
\usepackage{fancyvrb}
\usepackage{fvextra}
\hypersetup{
     colorlinks=true,
     linkcolor=black,
     filecolor=black,
     citecolor = black,      
     urlcolor=black}    
\usepackage{xcolor}
\usepackage{siunitx}
\usepackage[style=authoryear-comp,
            natbib=true,
            backend=bibtex,
            giveninits=true,
            maxcitenames=2,
            maxbibnames=99,
            ]{biblatex}

\renewbibmacro*{cite:labeldate+extradate}{%
  \printtext[bibhyperref]{%
    \iffieldundef{note}
      {\printfield{year}\printfield{extradate}} 
      {\printfield{note}
       \addslash
       \printfield{year}\printfield{extradate}}}}       
\renewbibmacro*{finentry}{%
  \finentry
  \iffieldundef{note}
    {}
    {\newunit\newblock{ (Original work published \printfield{note})}}
  }

\AtEveryBibitem{%
    \clearfield{month} 
    \clearfield{day}   
    \clearfield{url}
    \clearfield{isbn}
    \clearfield{issn}
    \clearfield{note}
    \clearfield{urldate}
    \clearfield{address}
    \clearfield{editor}
    \clearlist{language} 
}

\AtBeginBibliography{%
  \renewbibmacro*{urldate}{} 
}

\DeclareFieldFormat{addendum}{\mkbibparens{#1}\nopunct}
\DeclareFieldFormat{edition}{\mkbibparens{#1}}
\DeclareFieldFormat[inbook]{title}{#1} 
\DeclareFieldFormat[article]{title}{#1} 
\DeclareFieldFormat[book]{title}{\mkbibemph{#1}} 
\DeclareFieldFormat[book]{volume}{(Vol. #1)} 
\DeclareFieldFormat{journaltitle}{\mkbibemph{#1}} 
\DeclareFieldFormat[inproceedings]{title}{#1} 
\DeclareFieldFormat[book]{edition}{(#1). } 
\DeclareFieldFormat{pages}{#1} 
\renewbibmacro*{title}{%
  \printfield{title}%
  \iffieldundef{edition}{}{\setunit{\addspace}\nopunct}%
}

\renewbibmacro{in:}{}
\renewbibmacro*{in:}{%
  \ifboolexpr{
    test {\ifentrytype{inproceedings}} 
  }
  {\bibstring{in}\intitlepunct}  
  {}            
}
\renewcommand*{\intitlepunct}{\nopunct}

\renewbibmacro*{volume+number+eid}{
  \addcomma
  \textit{\printfield{volume}}%
  \iffieldundef{number}
    {}
    {\mkbibparens{\printfield{number}}}%
  \setunit{\addcomma\space}%
  \printfield{eid}
}

\usepackage[T1]{fontenc} 
\usepackage{mlmodern}
\usepackage[hang,flushmargin]{footmisc} 

\providecommand{\tightlist}{\setlength{\itemsep}{0pt}\setlength{\parskip}{0pt}}

\usepackage{etoolbox}
\AtBeginEnvironment{table}{\small}
\AtBeginEnvironment{tabular}{\small}
\AtBeginEnvironment{tabularx}{\small}
\AtBeginEnvironment{longtable}{\small}

\makeatletter
\renewcommand{\paragraph}{%
  \@startsection{paragraph}{4}{\z@}%
    {3.25ex \@plus1ex \@minus.2ex}%
    {-1em}%
    {\normalfont\fontsize{10}{12}\fontfamily{qhv}\selectfont\bfseries\protect\raisebox{-.225pt}}}
\makeatother

\makeatletter
\renewcommand{\maketitle}{
  \begin{flushleft}
    \fontfamily{qhv}\selectfont
    {\LARGE \textbf{\@title}} \\[2mm]
    \fontfamily{qhv}\selectfont \@author \\[2mm]
    \@date
  \end{flushleft}
}
\makeatother

\usepackage{caption}
\usepackage{subcaption}
\DeclareCaptionFormat{myformat}{#1#2#3\dotfill}


\AtBeginEnvironment{table}{%
  \small\usefont{T1}{qhv}{m}{n}\selectfont
}

\definecolor{darkgreen}{rgb}{0.0,0.5,0.0}
\definecolor{darkgray3}{rgb}{0.4,0.4,0.4}

\newcommand{\ntweets}{\num{17666756} }
\newcommand{\nspritzertweets}{\num{2186054} }

\newcommand{\percentageantagonization}{$32.7 \pm 2.5\%$}
\newcommand{\percentagecriticism}{$21.1 \pm 2.4\%$}
\newcommand{\percentageleftantagonization}{$28.4 \pm 4.0\%$}
\newcommand{\percentagerightantagonization}{$48.8 \pm 4.6\%$}
\newcommand{\percentageleftcriticism}{$23.8 \pm 4.0\%$}
\newcommand{\percentagerightcriticism}{$18.9 \pm 3.8\%$}

\newcommand{\percentagenaantagonization}{$18.9 \pm 4.5\%$}
\newcommand{\percentagenacriticism}{$20.1 \pm 4.4\%$}
\newcommand{\percentagenanone}{$61.0 \pm 4.6\%$}

\usepackage{graphicx}

\usepackage{booktabs}
\usepackage{diagbox}
\usepackage{multirow}
\usepackage{tabularx}
\usepackage{url}

\title{High prevalence of antagonizing rhetoric in trending\\[2mm] topics on Twitter/X}

\author{\vspace{1mm}\raggedright Armin Pournaki\textsuperscript{1,2,3,$*$}, Eckehard Olbrich\textsuperscript{2}}
\date{%
    \footnotesize
    \textsuperscript{1}Institute of Sociology, University of Bern, Bern, Switzerland\\%
    \textsuperscript{2}Max Planck Institute for Mathematics in the Sciences, Leipzig, Germany\\%
    \textsuperscript{3}médialab, SciencesPo, Paris, France\\[1mm]
    $^*$Corresponding author: armin.pournaki@unibe.ch\\[1ex]
    \today
}

\usepackage{silence}
\AtBeginEnvironment{table}{%
  \small\usefont{T1}{qhv}{m}{n}\selectfont
}
\AtBeginEnvironment{tabular}{%
  \small\usefont{T1}{qhv}{m}{n}\selectfont
}
\begin{document}
\allsectionsfont{\usefont{T1}{qhv}{b}{n}\selectfont}

\maketitle

\vspace{7mm}
\noindent
{\fontfamily{qhv}\selectfont \large \textbf{Abstract}}
{\small
\fontfamily{qhv}\selectfont \footnotesize  
\begin{spacing}{1.15}
Disagreement between political opponents is a central element of democracy. 
This disagreement can be voiced in different ways. 
An opponent can be antagonized, i.e. portrayed as an enemy that needs to be overcome, or they can be criticized, i.e. treated as a legitimate adversary with whom one can engage. 
While the latter is a fundamental part of democratic exchange, the former can be detrimental to public discourse. 
Political disagreement increasingly takes place on social media, where negativity towards political out-groups has been found to drive engagement. 
However, existing research has largely treated such negativity as a single dimension, leaving open whether it is antagonization or criticism that circulates on these platforms. 
This work addresses this gap by introducing a computational approach to detect these two types of rhetoric in text and identify its targets. 
On a corpus of over 17 million tweets from trending topics in the German Twittersphere, the method reveals that about a third of all tweets contain antagonizing rhetoric.
Comparing tweets from trending topics to a random sample shows that the discourse in trending topics is systematically more antagonizing than ordinary Twitter communication. 
Combining antagonization detection with a large-scale estimation of users' political leaning, we find that right-leaning users employ antagonizing rhetoric significantly more than their left-leaning counterparts. 
Furthermore, both forms of oppositional rhetoric are strongly amplified within political camps, with multipliers retweeting them up to twice as often as tweets without oppositional rhetoric. 
Finally, patterns of antagonized targets reveal how different political camps construct their out-groups, with the right-wing populist party AfD and the Green party as the central antagonists of the opposing camps. 
\end{spacing}
}
\vspace{2mm}

\section{Introduction}

Opposition is a central element of democratic political communication. Political debate entails that actors disagree and voice that disagreement. However, not all expressions of opposition emplot political opponents in the same way. On the one hand, opponents can be portrayed as legitimate adversaries, actors with whom one disagrees but whose place in political competition is acknowledged. On the other hand, they can be portrayed as antagonists, threats that have no legitimate place in the political arena and should be removed from it. Both voice opposition, and both can be harsh, or strategically deployed. The main distinction lies in whether the opponent's legitimacy is upheld or denied. We refer to the former mode of rhetoric as \textit{criticism} and to the latter as \textit{antagonization}. 

This distinction is important because the two modes relate differently to democratic discourse. Criticism, however forceful, implies that the opponent remains a participant in an ongoing exchange. Antagonization calls that status into question and may thereby erode the mutual respect and recognition necessary for deliberation and compromise. Antagonizing rhetoric may also shape group boundaries. From the perspective of group identity formation, a shared antagonist can serve as a central element for identification and demarcation, particularly when the in-group is heterogeneous \citep{Zehnter2025}. Antagonizing rhetoric is therefore one potential driver of the increasingly negative feelings toward out-groups that the literature refers to as \textit{affective polarization} \citep{Iyengar2012,Iyengar2015,Iyengar2019,Reiljan2020}, which has been associated with deepening social divisions and reduced support for democratic norms \citep{Kingzette2021}.

Social media platforms, in which users typically interact through short-form many-to-many communication, have been associated with such polarizing rhetoric \citep{Kumkar2025}. Prior work has identified out-group animosity in particular as a powerful driver of engagement online: polarizing posts referencing the political out-group are shared more often than those referencing the in-group, and out-group animosity is a stronger predictor of virality than negative affect or moral-emotional language \citep{Rathje2021,Yu2024,Ballard2023}. More generally, negative and high-arousal content is widespread in online political communication and tends to spread further than neutral or positive content \citep{Soroka2019,Brady2017,Watson2024}. Social media may thus offer favorable conditions for actors who seek to mobilize support through political conflict \citep{Engesser2017,Brady2021}.

So far, the literature around polarizing rhetoric on social media falls short in disentangling which form of political opposition it precisely measures. Existing large-scale analyses typically measure out-group negativity as a single dimension, and therefore cannot tell whether the prevalence and engagement of polarizing rhetoric reflect speech that denies opponents their legitimacy or (forceful) criticism that acknowledges it. Our work addresses this gap by systematically measuring the prevalence of both antagonization and criticism in a large corpus of over 17 million tweets from trending topics in Germany between 2021 and 2023. Trending topics capture the debates that attract the most public attention at a given time, and thus the arena in which polarizing rhetoric is most visible. Building on prior work that analyzed structural polarization and political narratives in polarized debates \citep{Pournaki2025}, we make three contributions. First, we introduce and operationalize a theory-driven distinction between antagonization, which portrays the political opponent as an enemy or threat without legitimate standing, and criticism, which contests an opponent's positions, actions, or ideas while acknowledging their legitimacy. Second, using an LLM-based annotation procedure combined with a statistical correction for classification error, we show that antagonization is widespread in German trending topics, and that right-leaning accounts antagonize substantially more than left-leaning ones. Third, our analysis of antagonized targets reveals that different political factions construct distinct out-groups, with the right-wing populist AfD and the Green party emerging as the main antagonists of the opposing sides. These findings contribute to better understanding the discursive dimension of affective polarization.

The paper proceeds as follows. Section~\ref{sec:theory} sets the theoretical background and relates our approach to theories of agonistic democracy and theories of collective identity formation. Section~\ref{sec:relatedwork} reviews related methodological work on computationally measuring affective polarization in text. Sections~\ref{sec:data} and \ref{sec:methods} present the data and methods employed to measure antagonizing and criticizing rhetoric on Twitter. Section \ref{sec:results} presents the main results: the estimated share of antagonizing/criticizing rhetoric in German Twitter trends, how it evolves over time, how it compares to a random sample, how it is amplified, in which topics it is most prevalent and who are its targets. Finally, Sec.~\ref{sec:discussion} discusses the implications of our findings, relating them to previous work and setting out the limitations of our approach. 

\section{Theoretical background}
\label{sec:theory}
\subsection{Defining antagonization}
\label{sec:definitions}
This work sets out to develop a definition that can be operationalized for the computational extraction of antagonizating rhetoric from large textual corpora. Political communication on social media is often characterized by a negative and adversarial tone \citep{Rathje2021}. However, not all negativity is equivalent. We draw an analytical distinction between two rhetorical modes of oppositional political discourse: on the one hand, expressing disagreement with ideas or policies while accepting the legitimacy of the political opponent; on the other, antagonizing the opponent personally by portraying them as a threat or enemy. This distinction carries different implications for democratic discourse. Where criticism preserves the possibility of deliberation with an adversary, antagonization implies the necessity of removing an enemy from the political landscape altogether. Based on these considerations, we propose the following definitions:

\textbf{Antagonization} is present when the author portrays a person, group, or entity as an enemy, threat, or hostile ``other,'' and/or employs contemptuous, demeaning personal attacks, including insults, ridicule, belittlement, or dehumanizing metaphors that delegitimize the target.

\textbf{Criticism} is present when the author expresses objections, disagreement, or negative evaluation without enemy framing or personal disparagement. In contrast to antagonization, the political opponent is acknowledged as a legitimate participant in the political process. The focus is on actions, decisions, policies, or ideas rather than on the person as such.

\subsection{Relations to the theory of agonistic pluralism}
\label{sec:mouffe}
The above distinction resonates with Chantal Mouffe's theorization of the political, in particular her distinction between antagonism, a friend/enemy relation in which the opponent's legitimacy is denied, and agonism, a relation between adversaries who contest each other's positions within a shared democratic framework \citep{Mouffe2000, Mouffe2005}. The concept of legitimacy is the clearest overlap in both frameworks: for Mouffe, the transformation of antagonism into agonism hinges on whether the opponent is recognized as a legitimate adversary. In our codebook, the same threshold is part of the distinction between antagonization and criticism. 

However, we note that our operationalization departs from Mouffe's theory in several respects. For one, we do not claim to operationalize her framework directly. Mouffe's antagonism is an ontological category describing the ever-present possibility of irreconcilable conflict in the political. Our categories operate at the discursive level, capturing how antagonizing or critical stances are expressed in text. An underlying antagonistic relation may exist without manifesting in the rhetoric our operationalization detects. 

Similarly, our category of criticism does not map directly onto Mouffe's agonism. For Mouffe, agonistic politics can be passionate, emotionally charged, and deeply contentious: what matters is not the heat of the discussion, but whether the adversary's right to participate is upheld. Our ``criticism'' category, with its emphasis on issue-focused and non-disparaging disagreement, captures only one register of what Mouffe would consider agonistic exchange. 

\subsection{Negational identities and chains of equivalence}
Mouffe's framework also points to a function of antagonization. Drawing on the post-structuralist notion of a ``constitutive outside'' \citep{Staten1984}, she argues that every collective identity requires a ``them'' from which an ``us'' can be distinguished, such that the antagonist becomes the condition for identity while at the same time being the threat that needs to be fought \citep{Mouffe2000}. This idea has an empirical counterpart in social psychology. While a group can be identified through the common features of its individuals, research in social psychology suggests an additional, potentially more powerful mechanism for group identification: \citet[p.794]{Zhong2008} define \textit{negational categorization} as ``the process by which an individual's identity is defined by outgroups, by what people are not (e.g., not liberals)''. The negation is especially useful when the in-group is not particularly homogeneous. By simply saying what the group is \textit{not}, one can circumvent the difficulties of uniting a group that has few common features, as has been shown to be effective in particular in right-wing populist parties \citep{Zehnter2025}.

The power of such negational identities has been demonstrated in various studies. \citet{Bosson2006} show that a negative attitude towards a third party is more effective in increasing closeness within a group. In the realm of political participation, \citet{Caruana2015} show that negative partisanship, i.e. an affective repulsion towards a given party can serve as a motivation for voting. \citet{Areal2022} show that the effects of negational identities can surpass in-group attachments, meaning that the negative affect towards ``them'' can be strong without a clearly defined ``us''. In political settings where the in-group is often difficult to define, it is thereby easier to unite the group through a common enemy. \citet{Renstrom2021} show, in the Swedish context, that the perception of an outside threat is positively related to intergroup affective polarization. Particularly through threat-based language, antagonizing speech therefore contributes to increasing affective polarization. 

While negational categorization describes a psychological mechanism, Mouffe's earlier work with Laclau describes how political actors can strategically employ it, in particular through their concept of ``chains of equivalence'' \citep{Laclau1985}. Their argument is the following: a variety of struggles exist in society that are not necessarily related to each other (e.g. climate change, social justice, gender equality). The political goal is to link those different struggles together such to consolidate and mobilize a potentially fragmented electorate. A possible link between these different struggles arises through common antagonists (neoliberalism, fossil fuel industry, patriarchy). As has been noted previously in the context of negational identities, it is useful to draw sharp borders excluding this antagonist from the in-group, especially if the in-group is diffuse. 
While Laclau and Mouffe developed this argument with emancipatory left-wing movements in mind, the mechanism itself is not bound to a particular ideology, and can equally serve to unite a heterogeneous right-wing electorate against common enemies.
For political entrepreneurs, it can therefore be beneficial to discursively antagonize in order to draw these boundaries delimiting the in-group from the antagonist in an explicit way, thereby connecting a variety of issues and groups together. In our data, we would accordingly expect antagonization to be directed not only at political opponents, but at a range of targets that together form the ``constitutive outside'' of an in-group. 

\section{Related methodological work}
\label{sec:relatedwork}
Previous work has focused on developing methods to measure affective polarization from raw text. We provide a brief overview of the different approaches that have been employed to achieve this before highlighting how our work departs by adding nuance to existing measurements. 

\paragraph{Dictionaries.} A part of the literature uses lists of words that are either adapted from existing sentiment dictionaries, or created in a domain-specific way to assess whether a given text contains traces of affective polarization. \citet{Roed2025} argue generic keyword lists miss political context and therefore build a domain-specific dictionary to capture how political elites speak about opposing party members. \citet{Garzon-Velandia2025} have gone further by developing and validating dictionaries meant specifically for negative affective polarization, including dimensions such as negative affect and delegitimization. While such domain-adapted dictionaries benefit from strong interpretability and domain vailidity, they may lack adaptability to other contexts, and it is difficult to capture the many varieties through which polarization can be expressed through a static dictionary. 

\paragraph{Sentiment and emotion.} Another set of studies use sentiment analysis tools to assess whether a given text is positive or negative, treating negative language toward political opponents as a sign of affective polarization. Such approaches have been employed to measure affective polarization in replies of non-likeminded users on Reddit \citep{Marchal2022} or members of parliament from opposing political camps \citep{Evkoski2025}. Other approaches go beyond binary sentiment and include emotions, such as anger or fear, as well as toxicity, highlighting systematic differences between in- and out-group replies on Twitter \citep{Lerman2024}. The shortcomings of sentiment analysis are that language can be more nuanced and that affective polarization can express itself through a variety of different ways that do not necessarily contain negative emotions (e.g. irony, sarcasm, ridicule). 

\paragraph{Supervised classifiers.} Another strand of work builds on training classifiers on labeled examples of affectively polarizing speech. Such approaches usually rely on annotated training data from which the model generalizes and can then target the construct of out-party negativity. Using this approach, \citet{Ballard2023} find that highly polarizing content in tweets from congressional members lead to more engagement. Similarly, \citet{Overgaard2024} develop a classifier to extract expressions of affective polarization in text and show how the topic of Covid-19 became polarized over time on social media. \citet{Fulay2026} classify political comments on Twitter and Reddit for positive or negative out-group sentiment and track rising animosity over time, showing that left and right-leaning communities show different levels of out-group animosity depending on the topic discussed. 
These supervised approaches, while adding nuance compared to sentiment-based approaches, typically operationalize out-party negativity as a single dimension, collapsing the distinction between substantive criticism and hostile antagonization that is central to our analysis. 

\paragraph{Large language models. }
More recently, large language models (LLMs) have been used for content analysis and proved to be quite successful \citep{Bermejo2025,LeMens2025}. This also extended to the task of identifying affective polarization. In their feed-based reranking experiment on X, \citet{Piccardi2025} use an LLM to classify partisan animosity in tweets in order to select those that should be downranked. \citet{Chowdhury2026} use LLMs and an approach based on heuristic scoring to quantify affective polarization from large-scale social media text, extracting stance, emotional tone, and agreement patterns and combining them to reveal distinct event-dependent patterns of anticipation and reaction-driven polarization. 

Methodologically, we go beyond existing approaches in several ways: first, our approach, leveraging LLMs, does not require any pre-defined keywords and can thus be transferred easily to a variety of domains. Second, we explicitly measure antagonization, which can be considered both a precursor of affective polarization (the enemy needs to be set in order for the group to develop animosity towards them), and a manifestation of existing animosity towards that antagonist. Third, we distinguish antagonization from criticism, separating opposition that denies the opponent's legitimacy from substantive disagreement, a distinction that measures of negativity or emotional tone cannot capture. Finally, our method simultaneously captures the target of antagonization, doing so without assuming pre-defined out-groups. Instead, the collectively constructed out-group emerges from an inductive analysis of the antagonized targets. 

\section{Data}
\label{sec:data}
The data used in this study consists of two Twitter collections. The first datset (D1) includes tweets from trending topics in Germany from April 2021 to July 2023 and consists of over 17 million \ntweets tweets.
The second dataset (D2), is a subset of the 1\% ``Spritzer'' dataset, covering three months in 2022 and one month in 2023, chosen because they overlap with our trending topic collection. To ensure comparability with the trending topic data, we keep only German-speaking tweets in the 1\% sample. This yields over 2 million tweets for the four months. We refer the reader to the Supplementary Material~\ref{sec:SMdata} for a detailed description of the data collection. 

\section{Methods}
\label{sec:methods}
\subsection{LLM-based extraction of antagonization}
For each tweet, we aimed to classify whether it expresses antagonization, criticism, or neither toward a specific target. To this end, we developed a codebook incorporating the definitions described in Sec.~\ref{sec:definitions} to manually annotate a random sample of tweets. 
The two authors, both native German speakers with domain knowledge in German politics, labeled 1000 tweets independently, yielding initial disagreement on 256 entries ($\kappa = 0.57$), which were subsequently reconciled in mutual discussion and agreement with the codebook to establish a ``ground truth'' dataset. See Supplementary Material ~\ref{sec:annotation_procedure} for details on disagreements between annotators.

We then developed a structured prompt based on the codebook and evaluated it across a range of open-weight models using \texttt{ollama}, which allows for rapid local deployment and testing. This initial screening revealed that models with a low parameter count (less than 7B) performed below acceptable thresholds for this task and were excluded from further consideration. For the remaining candidate models, we iteratively refined the prompt (reported in Sec.~\ref{sec:prompt}) to maximize alignment with human annotation. Based on this evaluation, \texttt{qwen3:32b} \citep{Yang2025} was selected and deployed via \texttt{vLLM} \citep{Kwon2023} for improved inference speed and throughput. The model achieves a macro F1 score of 0.7 and weighted F1 score of 0.75. See Supplementary Material \ref{sec:appendix:model-selection} for the systematic model comparison. 

Because the classifier is imperfect, applying it to the full corpus yields biased estimates of category proportions. We therefore report all category proportions using prediction-powered inference \citep{Angelopoulos2023}, which uses the human-annotated validation set to correct for classifier error and to construct confidence intervals that reflect this uncertainty. Full methodological details of the error correction are given in Supplementary Material~\ref{sec:appendix:ppi}.

The classifier also extracts the potential targets of antagonizing or criticizing rhetoric. Systematic comparison with the human annotation yields agreement in 84\% of the cases (see Supplementary Material~\ref{sec:appendix:target-evaluation} for details). In order to measure the prevalence of targets across political camps, we manually merge various instances of the same target. (e.g. ``Laschet'', ``Armin Laschet'', ``Kanzler Laschet''...) The procedure and the thusly merged targets are described in the Supplementary Material~\ref{sec:target_merges}.

\subsection{Structural analysis of issue polarization}
The estimation of ideological positions of users emitting the tweets was previously published in \citet{Pournaki2025} and will be summarized here. This previous analysis allows to analyze patterns of antagonizing rhetoric across issues, through topic modelling, and the ideological camps, through the network-structural analysis of retweet networks. 
\paragraph{Extracting issues from tweets}
\label{sec:topicmodel}
To identify the main issues discussed across trends, all tweets were passed through a topic modeling pipeline based on BERTopic \citep{Grootendorst2022}. This approach leverages pre-trained sentence embeddings to project tweets into a semantic space, where clustering algorithms inductively surface latent topics. The underlying intuition is that geometric proximity in this space reflects thematic similarity. The resulting fine-grained topics were subsequently aggregated by hand into broader issues: for instance, Covid-related topics (e.g., vaccination, mask mandates, school closures) or climate change (e.g., activism, heatwaves). Each trend was then labeled with the issue category corresponding to the majority of its constituent tweets. A full description of the extracted topics is provided in \citep{Pournaki2025}.
\paragraph{Extracting opinions from tweets}
To capture the opinion structure within each trend, a retweet network was constructed for every trending topic, in which nodes represent Twitter accounts and a directed edge from account $i$ to account $j$ indicates that $i$ retweeted $j$. Building on the established interpretation of retweets as signals of endorsement, densely connected communities in such networks are understood to reflect shared opinion stances in the underlying political discourse \citep{Conover2011a,Gaumont2018,Gaisbauer2021}. A trend's retweet network can thus be treated analogously to an opinion poll, with each user assigned a stance of $-1$ or $1$. Community detection was performed using a Stochastic Block Model (SBM) \citep{Peixoto2019}, yielding a cluster membership for each account in each network. Aggregating these assignments across the full dataset allowed to quantify how frequently any two users co-occur in the same opinion cluster. A user alignment score derived from this co-occurrence frequency reveals a broad two-camp structure in the user base (one left-leaning and one right-leaning). The leaning was identified through manual inspection of the most active accounts in each group. We focus the analysis on two types of power users that are defined by their (re)tweet behavior: influencers are the 1000 users that get retweeted the most (and therefore have the highest in-degree across trending topics), while multipliers are the 1000 users that retweet the most (highest out-degree), and act as intermediaries on the platform. 
\paragraph{Assigning political leaning to tweets}
The global clustering into left/right is used to assign each cluster of individual trend-based retweet networks to one of the two global clusters, based on the majority leaning of its users. Those clusters that have no users from either global left/right cluster are labeled as ``not assigned''. Consequently, each tweet and retweet in the dataset is labeled as one of either \{left, right, not assigned\}, based on the cluster of the user that emitted the (re)tweet: if an account from a left-leaning cluster retweets an antagonizing tweet from the right-leaning cluster, this retweet event is added to the overall antagonization count of the left-leaning cluster. 
%
\section{Results}
\label{sec:results}
\subsection{Antagonization statistics}
\begin{figure}[t]
    \includegraphics[width=\linewidth]{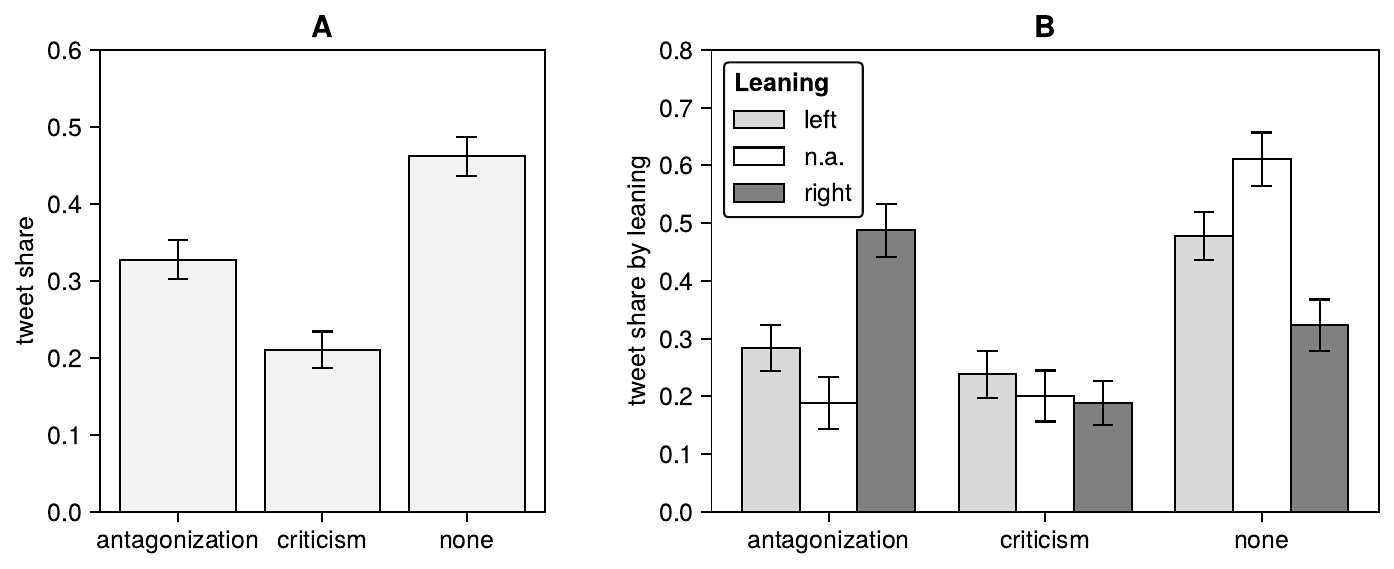}
    \caption{\textbf{Estimated shares of antagonization, criticism and none-labeled tweets across the dataset (A) and divided by political leaning of authors (B).} Bars show prediction-powered point estimates. Error bars represent 95\% prediction-powered confidence intervals.}
    \label{fig:antagonization_share_global}
\end{figure}
\begin{figure}[t]
    \includegraphics[width=\linewidth]{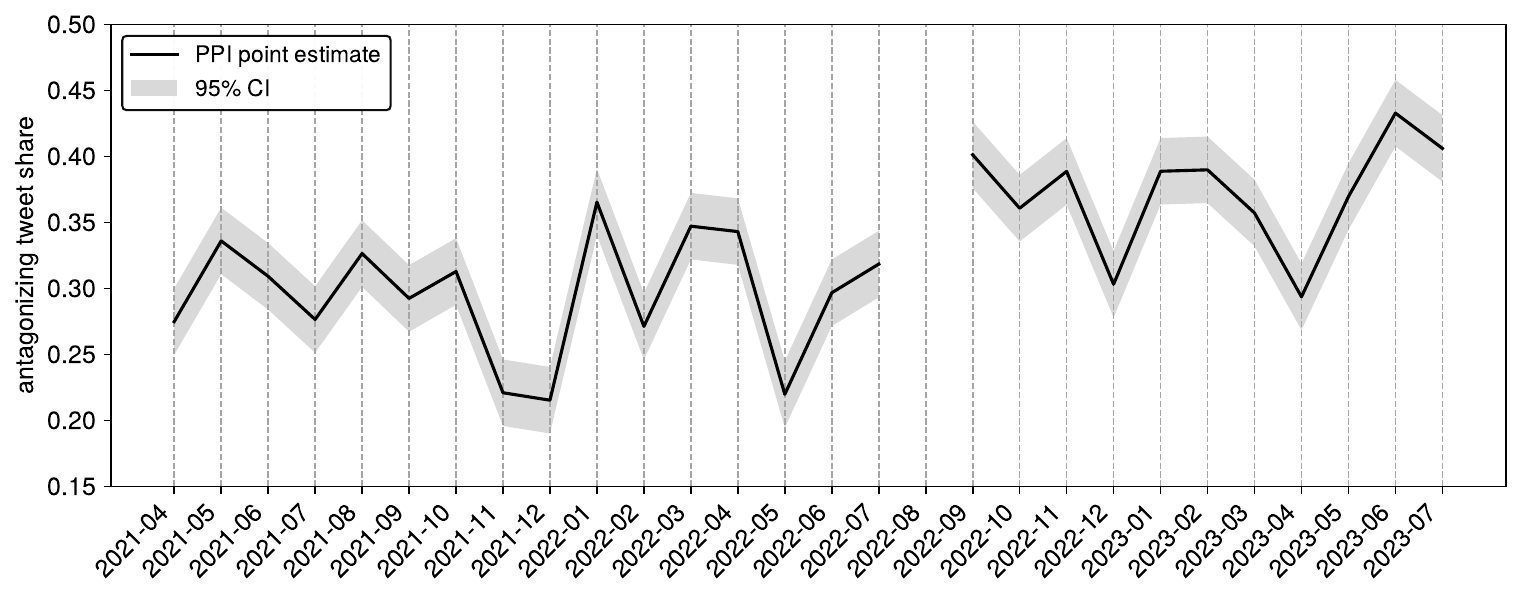}
    \caption{\textbf{Share of antagonizing tweets over time.} The line shows the monthly prediction-powered estimate of the share of antagonizing tweets, the shaded band the pointwise 95\% confidence interval. After September 2022, the share of antagonizing tweets is higher than in the preceding period, except in December 2022 and April 2023, when it drops back to levels typical of the earlier period. No data are available for August 2022.}    
    \label{fig:excess_antagonization}
\end{figure}
\begin{figure}[t]
    \includegraphics[width=\linewidth]{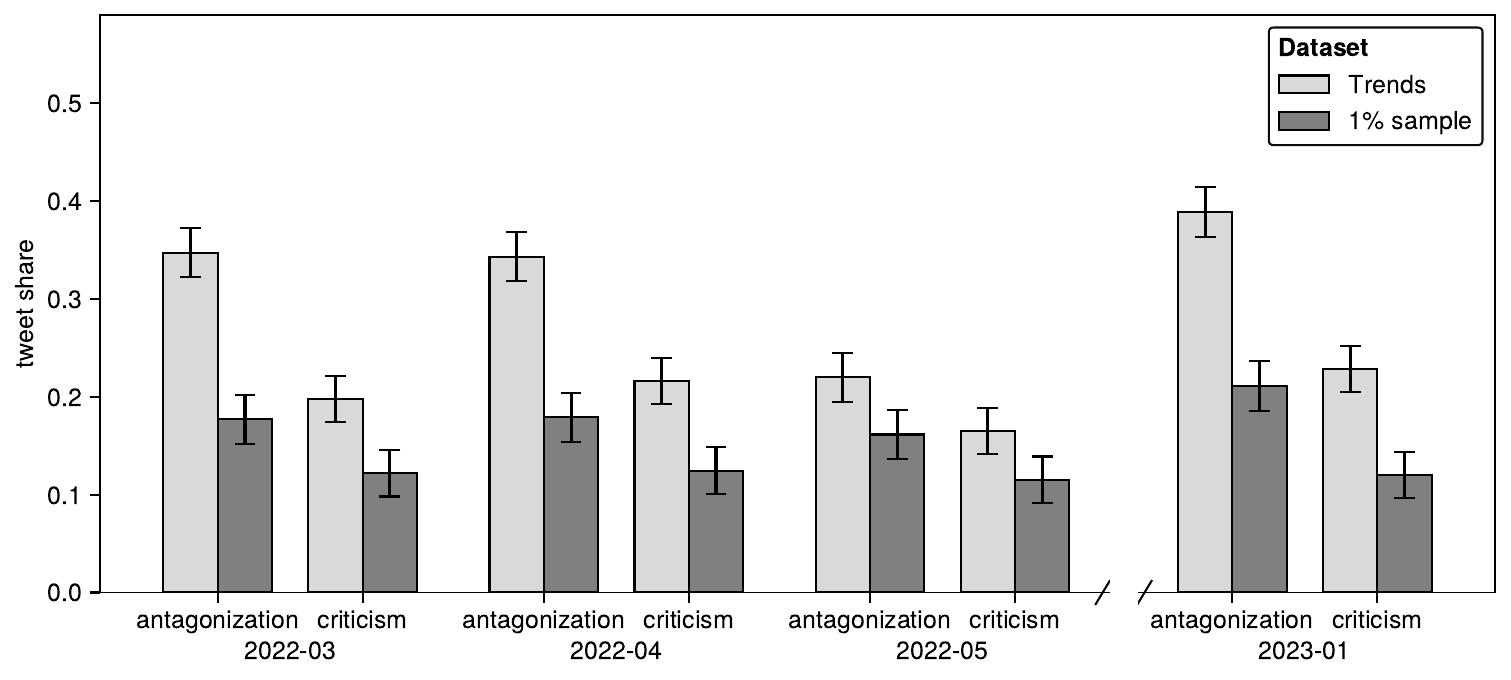}
    \caption{\textbf{Monthly comparison with a 1\% random sample.} Shares of antagonizing and criticizing tweets in trending topics and in a 1\% random sample of tweets, for the four months in which both datasets are available (March, April and May 2022, and January 2023). Error bars show 95\% confidence intervals of the prediction-powered estimates. In every month, trending topics contain more antagonization and more criticism than the random sample. In both datasets, the share of antagonization is higher in January 2023 than in the months of 2022.}    
    \label{fig:random_sample_comparison}
\end{figure}
\paragraph{Antagonization in German trending topics.} How prevalent is antagonizing speech in tweets from trending topics? Figure~\ref{fig:antagonization_share_global} (A) shows the share of tweets that contain antagonizing or criticizing content across the dataset. We report that \percentageantagonization~of tweets collected from trending topics contain antagonization, while \percentagecriticism~contain criticism. The remainder of tweets are categorized as ``none''. 
Looking at the share of antagonizing tweets by political cluster in Fig.~\ref{fig:antagonization_share_global} (B), we observe an ideological asymmetry with respect to the amount of antagonizing content: \percentagerightantagonization~of right-leaning tweets contain antagonizing rhetoric, compared to \percentageleftantagonization~of tweets from the left-leaning camp. For criticism, the asymmetric effect is reversed, but not as pronounced (\percentageleftcriticism~on the left vs \percentagerightcriticism~on the right). 
The tweets that could not be assigned to a political cluster (n.a.) predominantly fall into the ``none'' category (\percentagenanone~none, \percentagenaantagonization~antagonization and \percentagenacriticism~criticism). This is likely because these accounts may participate less in political discussions, such that a larger share of their tweets does not concern political opponents. 

In sum, we state that antagonizing rhetoric is prevalent in trending topics on Twitter/X, while being more pronounced in the right-leaning community than the left-leaning counterpart. 


\begin{figure}[t]
    \includegraphics[width=\linewidth]{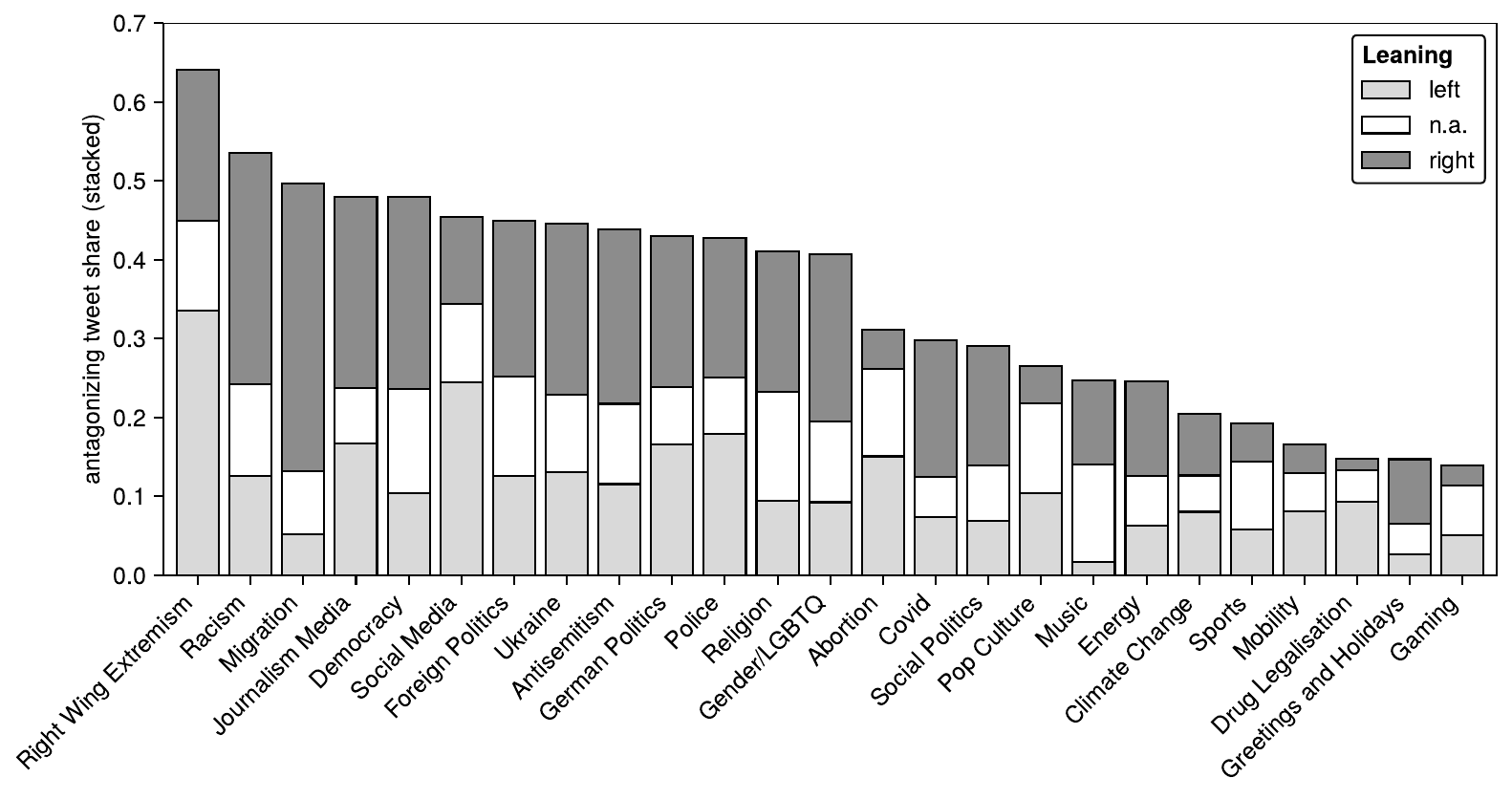}
    \caption{\textbf{Antagonization by issue.} For each issue, we show the share of antagonizing tweets, broken down by the political leaning of the authors. For several issues, the contributions of the two camps differ markedly, for example \textit{Right-Wing Extremism}, where antagonization comes mainly from left-leaning accounts, and \textit{Migration}, where it comes mainly from right-leaning accounts.}    
    \label{fig:antagonization_by_topic}
\end{figure}
\paragraph{Antagonization over time.} Figure~\ref{fig:excess_antagonization} shows the monthly share of tweets that cointain antagonizing rhetoric. Comparing the mean share of antagonization between two equal time periods $t_1= $ [2021-10 -- 2022-07] and $t_2= $ [2022-10 -- 2023-07], we observe a rise from $28\pm 2.5\%$ to $37 \pm 2.5\%$. 
It is worth noting that certain months exhibit relatively low shares of antagonization, which we relate to that month's most prevalent trending issues (see Supplementary Material Fig.~\ref{fig:topics_over_time} for a detailed account of monthly issues). For November 2021, we find that \textit{Covid} was most prevalent, which contained a lower share of antagonizing tweets in our dataset due to the prevalence of informational content (see Fig.~\ref{fig:antagonization_by_topic}). In December 2022, the main issue revolves around Christmas in \textit{Greetings and Holidays}, and in May 2022, the most salient issue is \textit{Sports}, which usually usually do not trigger (politically) antagonizing rhetoric. 

\paragraph{Comparison with a random sample.}
Is antagonizing language overrepresented in tweets from trending topics? To answer this question, we analyzed tweets from the 1\% ``Spritzer'' dataset for four months that overlapped with our collection. Figure~\ref{fig:random_sample_comparison} shows the comparison of antagonization and criticism shares in our dataset with this 1\% random sample of tweets for March, April, May 2022 as well as January 2023. We observe that (i) antagonization is prevalent in the random baseline, where on average 18.2\% of the tweets are antagonizing (between 16.1 and 21.1\% across months), (ii) it is also more prevalent than criticism (12.0\% on average), (iii) the shares of antagonization and criticism are higher in trending topics than in the random baseline in every month, with trending topics containing on average almost twice as much antagonization (32.5 vs.\ 18.2\%) and criticism (20.2 vs.\ 12.0\%), and (iv) the share of antagonizing content rises in both our dataset and the random sample between 2022 and 2023, from 30.3 to 38.9\% in trending topics and from 17.2 to 21.1\% in the random sample, comparing March to May 2022 with January 2023.
\paragraph{Antagonization by issue.} Do certain issues trigger more antagonizing speech more than others? Figure~\ref{fig:antagonization_by_topic} shows the share of antagonizing tweets per issue. The most antagonizing issue is \textit{Right Wing Extremism} with 64\%, followed by \textit{Racism} (54\%) and \textit{Migration} (50\%). The least antagonizing set of issues are \textit{Gaming}, \textit{Greetings and Holidays} and \textit{Drug Legalisation} with roughly 15\%. However, even within those seemingly apolitical issues, we observe examples of antagonization that relate to the larger ideological divides between the camps, such as antagonizing language against trans people on Mother's day in \textit{Greetings and Holidays}, or misogynistic delegitimization of female gamers within the \textit{Gaming} topic. 

It may at first glance be surprising to see that a polarizing issue such as \textit{Covid} is relatively less antagonizing (about 30\% of antagonizing tweets vs. for example 41\% in \textit{Gender/LGBTQ}). A manual inspection of the 100 most retweeted tweets that were not labeled as either antagonization or criticism reveals that the discourse is dominated by calls for action (vaccination, wearing masks) and information-based content (incidence numbers in various countries, ICU occupancies...) that did not contain any targets of criticism or antagonization. 

Looking at the way the left and right-leaning groups employ antagonizing rhetoric across issues, we observe stark issue-wise differences, highlighting that the camps do not contribute equally to the antagonization of out-groups. The most pronounced example is \textit{Migration}, where more than 73\% of the antagonizing content can be attributed to accounts from the right-leaning camp. Other asymmetries can be observed that point towards contentious topics in which the antagonist plays a central role: the issues of \textit{Right-Wing Extremism}, \textit{Abortion}, \textit{Social Media} in which the left predominantly antagonizes, and \textit{Racism}, \textit{Journalism/Media}, \textit{Covid}, \textit{Energy} on the right. 
\subsection{Amplification of antagonizing and criticizing rhetoric}
\begin{table}[htbp]
\caption{\textbf{Amplification of antagonization and criticism.} Share of original tweets and mean amplification factors by source and amplifier group, where amplification factors are relative to tweets without oppositional rhetoric.}
\label{tab:amplification}
\begin{tabular}{llccccc}
\toprule
 & \multicolumn{3}{l}{\textbf{Antagonization}} & \multicolumn{3}{l}{\textbf{Criticism}} \\
\cmidrule(lr){2-4} \cmidrule(lr){5-7}
 & Original & \multicolumn{2}{l}{Amplification by} & & \multicolumn{2}{l}{Amplification by} \\
\cmidrule(lr){3-4} \cmidrule(lr){6-7}
Source & tweet share & Left multiplier & Right multiplier & tweet share & Left multiplier & Right  multiplier\\
\midrule
Left influencer  & 22.0~$\pm$~4.0\% & 2.06 & 1.07 & 16\% & 2.17 & 1.51 \\[\medskipamount]
Right influencer & 44.0~$~\pm$4.6\% & 0.82 & 1.79 & 17\% & 1.26 & 1.83 \\
\bottomrule
\end{tabular}
\end{table}
Does antagonizing and criticizing rhetoric systematically yield more retweets? To answer this question, we first compute a baseline defined as the average number of retweets of a ``none''-categorized tweet, $t_\mathrm{none}$. The amplification factor $a_i$ of category $i$ is then given by $t_i/t_\mathrm{none}$, where $t_i$ is the average number of retweets a tweet from category $i$ recieved. When $a_i > 1$, tweets of $i$ are amplified. This computation yields an overall amplification factor of 1.62 for antagonization and 2.18 for criticism. 

Going further, we compute the amplification factors group-wise, for influencers and multipliers of the two opposing camps (see Tab.~\ref{tab:amplification}). Multipliers of both camps amplify antagonizing rhetoric of their own group quite strongly: left-leaning multipliers retweet an antagonizing tweet by left-leaning influencers about twice as often as they would retweet a ``none''-categorized tweet from that same group (2.06). The effect is a bit weaker for the right-leaning camp (1.79). Criticism from their own camp is amplified about equally strongly (2.17 for left-leaning and 1.83 for right-leaning multipliers).

Looking at cross-camp retweet interactions, we observe that multipliers amplify criticism coming from the opposing camp more strongly than antagonization: right-leaning multipliers amplify criticism by left-leaning influencers by a factor of 1.51, compared to 1.07 for antagonization, and left-leaning multipliers amplify criticism by right-leaning influencers by a factor of 1.26, while antagonization from the right-leaning camp is retweeted less often than ``none''-categorized tweets (0.82). We note however that these amplification factors are computed on relatively small numbers since multipliers retweet influencers of the opposing camp extremely rarely (<1\% on the left and 3\% on the right).

Note also the stark difference in prevalence of antagonizing rhetoric in original tweets of left- vs.\ right-leaning influencers. While $44\pm4.6\%$ of right-leaning influencers' tweets contain antagonizing rhetoric, only $22\pm4.0\%$ of left-leaning ones contain it. The right-leaning camp thus supplies considerably more antagonizing content, while the relative amplification of this content by multipliers is slightly weaker than in the left-leaning camp.

\subsection{Target analysis}
\begin{table}[t]
\caption{\textbf{Most antagonized targets by political camp.} The 20 most frequent targets of antagonizing tweets for left-leaning, right-leaning and unassigned accounts, after merging different references to the same target (see Sec.~\ref{sec:target_merges}). Target names are translated from German by the authors.}
\label{tab:targets_antagonization}
\begin{tabular}{l@{\hspace{25pt}}lr@{\hspace{50pt}}lr}
\toprule
 & \multicolumn{2}{c}{LEFT} & \multicolumn{2}{c}{RIGHT} \\
 & target & count & target & count \\
\midrule
1 & AfD & 142480 & Green Party & 142642 \\
2 & Armin Laschet & 112848 & Annalena Baerbock & 123019\\
3 & CDU & 87841 & Karl Lauterbach & 100537 \\
4 & Querdenker & 54720 & Robert Habeck & 65973 \\
5 & Vladimir Putin & 50076 & Media & 61572 \\
6 & Russia(ns) & 40615 & AfD & 54604 \\
7 & FDP & 40290 & Government & 53820 \\
8 & Sahra Wagenknecht & 38797 & Russia(ns) & 43226 \\
9 & Nazis & 33071 & Olaf Scholz & 42843 \\
10 & Friedrich Merz & 30651 & Migrants/refugees & 38589 \\
11 & CDU/CSU & 28020 & Politicians & 36273 \\
12 & Green Party & 27730 & Unvaccinated people & 33695 \\
13 & Olaf Scholz & 25472 & Traffic light coalition & 29207 \\
14 & Christian Lindner & 25254 & SPD & 27845 \\
15 & Hans-Georg Maaßen & 23371 & Angela Merkel & 26498 \\
16 & Annalena Baerbock & 22387 & Germany & 25570 \\
17 & Police & 21720 & CDU & 25447 \\
18 & Jens Spahn & 20937 & FDP & 22271 \\
19 & SPD & 20532 & they/you & 21083 \\
20 & CSU & 18158 & Tagesschau & 19087 \\
\bottomrule
\end{tabular}
\end{table}

\paragraph{Main targets of antagonization.} Who are the main antagonists of the opposing political camps? Table \ref{tab:targets_antagonization} shows the 20 most antagonized targets for left- and right-leaning accounts. This reveals the political left/right divide, with the Green party/ the AfD as anchors and top antagonized targets. Further, the left strongly antagonizes the Querdenker movement that promoted anti-vaccine messages during Covid, as well as Vladimir Putin for the invasion of Ukraine. In the right-leaning camp, we see the ``media'' as a central antagonist, as well as the government. The latter observation reflects the right-leaning camp's role of opposition during the observed time-period. Further down, we see migrants and refugees as salient targets of right-leaning antagonization, reflecting common right-wing populist narratives. 

Comparing the 20 most frequent targets of antagonization to those of criticism (see Fig.~\ref{tab:targets_criticism}), we do not observe any systematic discrepancies, suggesting that antagonization and criticism are used similarly for the main targets of both camps. Two notable exceptions arise: i) the AfD, which is the most antagonized target on the left, is not among the top 20 most criticized targets, and ii) the same holds for ``Russia(ns)''. 

\paragraph{Self-victimization}
A notable finding is the appearance of the AfD among the most frequently targeted entities within the right-leaning camp (rank 6). Manual inspection of a subset of thusly classified tweets suggests that this is partly driven by self-victimization narratives or the strategic weaponization of antagonization, whereby users invoke attacks attributed to the opposing camp in order to discredit it. 
One illustrative example, translated by the authors, reads: ``The Office for the Protection of the Constitution is responsible for keeping the AfD small. That’s how democracy is supposed to work — the domestic intelligence agency ensures the right election results.''
While such instances of self-victimization differ in their prevalence across camps, genuine intra-camp antagonization toward these targets is also present, such as heavy critique of the Green party on the left for neglecting their climate change goals once they were in government. 



\section{Discussion}
\label{sec:discussion}
We analyzed patterns of oppositional political rhetoric on social media using a large corpus of tweets from German trending topics between 2021 and 2023.
We introduced a theoretically motivated distinction between antagonization, where the opponent's legitimacy is questioned, and criticism, where the opponent is considered a legitimate contender in political competition. 
Using an LLM-based classifier, we showed that antagonization is strongly prevalent in trending-topic discussions, with more than 30\% of tweets containing antagonizing rhetoric, compared to about 20\% that contained criticizing content. Measuring the contributions of politically distinct camps, we found that right-leaning accounts use antagonizing rhetoric at significantly higher rates than left-leaning ones (48.8\% vs.\ 28.4\%).
Compared with a random sample of tweets from the same period, trending topics contain nearly twice as much antagonizing rhetoric. Over time, the share of antagonizing content increases. It rises from 28\% in October 2021 to July 2022 to 37\% in the same months one year later. 
Looking at the most active users reveals that multipliers, i.e. accounts that systematically and extensively retweet content, act as funnels of antagonization and criticism, selectively amplifying this type of rhetoric and thereby reinforcing their role as intermediaries in the diffusion of oppositional political discourse. This phenomenon is present in both left and right-leaning camps. 
Finally, an examination of targets of antagonization reveals how the two political camps construct their out-groups, with the central antagonists consisting of the Green party and the AfD. 

The high prevalence of antagonization we observe is consistent with prior findings on the negativity of political communication on social media \citep{Rathje2021,Soroka2019,Brady2017}, but our contribution extends this work by disentangling the various facets of negativity. Existing measures, from sentiment analysis to supervised classifiers of out-party negativity, conflate substantive policy disagreement with personal hostility. Our framework distinguishes antagonization, which portrays the opponent as an illegitimate threat, from criticism, which predominantly targets actions and policies while importantly preserving the adversary's legitimacy. 

This distinction is important because antagonizing rhetoric, aiming at excluding political opponents from the discourse can reflect anti-pluralistic tendencies. A central norm of democratic competition is the mutual toleration of political rivals, i.e., the acceptance of democratic opponents as legitimate competitors who have an equal right to exist and to govern \citep{Levitsky2018}. Antagonization violates this norm. If political opponents are portrayed as enemies or threats, their participation in political competition, and ultimately their electoral success, also appears illegitimate which undermines the willingness to accept electoral results on which democratic stability depends \citep{Anderson2005}. 

Antagonizing rhetoric may also contribute to affective polarization: repeated casting of political opponents as enemies may lead to negative feelings towards those adversaries. 
At the same time, those who are antagonized may respond with animosity towards the antagonizing side, such that antagonization can reinforce itself on both sides. 
In our data, we additionally observe instances in which being antagonized is itself used rhetorically. Users point to antagonization by the other side, portraying their own group as its victim, in order to spur negative feelings towards the antagonizers. 
Beyond being a potential driver, antagonizing rhetoric may also be an expression of existing animosity and thus an indicator for affective polarization. 
Affective polarization, in turn, has also been associated with lower support for democratic norms such as constitutional protections \citep{Kingzette2021}. 

Furthermore, antagonization may serve to sharpen group boundaries. Following \citet{Coser1956}, common enemies strengthen the cohesion of in-groups, which may be particularly relevant for populist rhetoric, where the in-group is often unclearly defined and diffuse \citep{Zehnter2025}. 
Our amplification results are consistent with this function. While criticism is amplified across political camps (with factors of 1.51 and 1.26), antagonization barely crosses camp boundaries (1.07 and 0.82), although these cross-camp factors rest on relatively few retweets. Antagonizing rhetoric thus appears to circulate mainly within the camp in which it is produced, which suggests that it serves internal cohesion rather than persuasion of the other side.

Measuring the prevalence of antagonizing rhetoric across political camps, we observed a stark asymmetry: \percentagerightantagonization~of right-leaning tweets are antagonizing, while \percentageleftantagonization~of tweets from the left-leaning camp are antagonizing (for \textit{influencers}, the most retweeted users, this is even more pronounced: 44\% on the right vs 22\% on the left). How can this be explained? Related work measuring the prevalence of incivility in congressional tweets identified a party's role in the opposition as a strong predictor for incivility \citep{Ballard2022,Tornberg2026a}. This may partly explain the prevalence of right-leaning antagonization compared to left-leaning, since the government in Germany during the majority of the observed period consisted of a coalition between the Social Democratic Party, Green Party and Free Democratic Party, containing two left-leaning parties, with the AfD and CDU/CSU in the role of opposition. 

A closer look at the targets of antagonization offers an additional perspective on this asymmetry. Besides political opponents, the right-leaning camp frequently antagonizes the mainstream media and migrants. This pattern is characteristic of right-wing populist communication, which typically relies on the construction of multiple enemies, not only political elites, but also institutions such as the mainstream media and out-groups such as migrants \citep{Mudde2004,Mudde2013,Kramer2021}. The targets we observe are therefore consistent with populist strategies. 

Turning our attention to the different types of users that contribute to these trends, we find that antagonization and criticism are strongly amplified by \textit{multipliers}, users that predominantly retweet content. Such users have previously been identified as intermediaries that act as funnels of ideologically coherent content for their followers \citep{Pournaki2025,Green2025}.
Compared to the average number of times they retweet an unclassified tweet, multipliers in our dataset retweet antagonization from their own camp's influencers about twice as much. For criticism, this amplification is even slightly stronger. Here, the effect of ideological position is reversed: left-leaning multipliers amplify antagonizing content of left-leaning influencers more than their right-leaning counterparts (2.06 vs 1.79). 

Looking at the overall amplification factors of antagonizing and criticizing rhetoric across the whole userbase, we notice that criticism is more strongly amplified than antagonization (2.19 vs 1.62). There is a feedback loop between user behaviour and algorithmic amplification... We cannot disentangle the two but formulate possible explanations for the patterns we observe. 
For one, the reputational risk of retweeting antagonizing content is higher than for criticism, which is more broadly accepted. Similarly, criticism may be more amplified across political camps because it does not polarize as strongly at antagonization. Looking at the cross-camp amplification patterns, even though cross-camp retweeting is rare, supports this hypothesis: the amplification factors across camps are higher for criticism than for antagonization.

Comparing our findings to a 1\% random sample of tweets, available for four months of the observed period, allows us to draw several additional conclusions about the discourse that circulates on the platform. Firstly, we observe that even in the random sample, antagonizing rhetoric is more prevalent than criticism, although the difference is less pronounced (18.2 vs.\ 12\%, averaged over the four months). Since the sample contains all sorts of tweets regardless of whether they were seen or retweeted, this suggests that, already at the supply level, social media users produce more antagonizing than criticizing content. Secondly, we highlight that trending topics contain much more antagonizing content than this random sample (32.5 vs.\ 18.2\%).
This is likely due to the content of trending topics. While the random sample contains a large share of non-political everyday communication, trending topics concentrate the most visible, and often contested, political debates of a given moment, in which opposition and conflict are particularly prevalent.

Finally, focusing our attention on the temporal evolution of the observed patterns, we see a rise in antagonizing rhetoric after September 2022, both in the trending-topic dataset (28.4\% from October 2021 to July 2022 vs.\ 37.2\% in the same months one year later) and in the random sample (17.2\% from March to May 2022 vs.\ 21.1\% in January 2023).
This could point towards a broader shift on the platform. One possible explanation is the change in the platform's ownership: Elon Musk's acquisition of Twitter, announced in April 2022 and completed in late October 2022, was followed by reduced content moderation, the reinstatement of previously suspended accounts, and changes to verification and ranking. Prior work documents a marked rise in hate speech on the platform in this period, beginning shortly before the acquisition was completed \citep{Hickey2023}. The timing in our data, however, is only partly consistent with this explanation, as the increase is already visible in September 2022, before most of these changes took effect. 

Our findings should be read in light of several limitations which invite for future research. 
First, our measurements rely on an LLM-based classifier, which is not error-free (macro F1 of 0.70). We account for this by reporting prediction-powered estimates, which use our human annotation to correct for the systematic errors of the classifier. These corrections rest the assumption that the errors of the classifier are stable over the observed period, and that they transfer to the random sample, where no separate annotation was conducted. 

Second, the ``criticism'' category is difficult to identify for both humans and LLM-based classifiers. 
The disagreements between annotators in the initial labeling round illustrate this, even in the presence of a detailed codebook. In addition, both the classifier and the annotators judged individual tweets in isolation. The meaning of a tweet may depend on the conversation it is part of, on linked content, on irony, or on references to earlier events, and antagonization in particular may be constructed across many tweets without any single one doing so explicitly. 

Third, we could ask what other discursive strategies may characterize political communication aside from antagonization and criticism. About 50\% of all classified tweets contained either one of the strategies. There could be others, especially on the left leaning side of the spectrum, where antagonization makes up about 28\% of tweets. Previous work on narrative patterns in polarized debates has identified solidarity as a meta-narrative that is particularly prominent on the left \citep{Pournaki2025a}. The computational identification of solidarity rhetoric, as a potential counterpart to antagonization for promoting in-group cohesion, represents a promising avenue for future work. 

Fourth, while our framework is motivated by the different implications of criticism and antagonization for democratic discourse, it does not imply that antagonization is always harmful. Denying legitimacy to an actor may be justified when that actor itself seeks to undermine democratic principles. Similarly, social movements that call out injustice may rely on sharp boundaries between those affected and those responsible, for instance to make structural injustice visible or to hold powerful actors to account. Our classification does not make these normative distinctions, and disentangling different forms of antagonization, according to their targets and their justification, remains an important task for future work.

Lastly, several of the open questions raised by our analysis could be addressed in future work. Comparing antagonization across platforms would show whether the patterns we observe are specific to Twitter/X. Following the same users before and after September 2022 would allow to separate changes in user composition from changes in behavior, and analyses at the level of conversations rather than single tweets could capture antagonization that is constructed across many tweets. Linking antagonizing rhetoric to survey measures of affective polarization would allow us to test the mechanism we propose more directly. Such analyses depend on access to platform data, which has become considerably more restricted since 2023. The provisions for researcher data access in Article 40 of the EU Digital Services Act \citep{DSA2022} may open up new possibilities in this regard. 

Despite these limitations, our study provides a systematic account of how political opposition is expressed on social media. Building on a theory-driven distinction between antagonization, which denies the opponent's legitimacy, and criticism, which contests positions and actions while acknowledging it, we find that opposition is highly prevalent in German trending topics: more than half of all tweets contain either antagonizing or criticizing rhetoric. Disentangling the two reveals that antagonization is more common, accounting for about a third of all tweets, and that it is considerably more prevalent among right-leaning than among left-leaning accounts. An analysis of the antagonized targets further shows that the two political camps construct distinct out-groups, with the AfD and the Green party as their central antagonists. These results show that what would appear as uniform negativity consists of two modes of opposition with different implications for democratic discourse. Separating them is therefore a necessary step towards better understanding the discursive dimension of affective polarization.

\renewcommand\bibname{References}
\printbibliography[heading=bibintoc]

\section*{Funding statement}
AP and EO acknowledge funding by the European Union’s Horizon Europe program under grant agreement ID 101094752: Social Media for Democracy (SoMe4Dem) — Understanding the Causal Mechanisms of Digital Citizenship.

\clearpage

\appendix
\renewcommand{\thesection}{S.\arabic{section}}
\renewcommand{\thesubsection}{S.\arabic{section}.\arabic{subsection}}
\renewcommand{\thefigure}{S\arabic{figure}}
\setcounter{figure}{0}
\renewcommand{\thetable}{S\arabic{table}}
\setcounter{table}{0}

\section*{\LARGE Supplementary Material}
\section{Data}
\label{sec:SMdata}
\subsection{Trending tweets in Germany}
The dataset includes tweets on trending topics over the course of more than two years. More specifically, we collected tweets from 2021-03-29 to 2023-07-12 according to the following scheme: at the beginning of each day, we launched a script that collects the current ``trending topics'' (from now on referred to as ``trends'') in Germany using the Twitter Trend API (v1). By default, trends are personalized based on the account's Twitter/X usage. One can, however, disable the personalization by setting a specific location from which to draw the trending topics, which then yields ``popular topics among people in a specific geographic location'' \citep{X/Twitter2025}. We re-ran the script every 15 minutes. At the end of each day, we counted the number of times each trending topic appeared during the day and kept the top 5 most frequent ones. This gave us a proxy of the five most important trending topics for that day. We then used the Twitter Search API (v1) to collect German-speaking tweets using the exact trend keyword as a query on the day it trended and the day after (48hrs). All the tweets were collected using a single Twitter API key, collecting tweets for maximally 24 hours every day. This yields a total of 19,105,532 across 2693 trends. 
\subsection{1\% random sample}
To compare the prevalence of antagonization in trending topics with a random baseline of platform activity, we rely on the Twitter ``Spritzer'' stream, a uniform 1\% random sample of all public tweets delivered in real time through Twitter's public Streaming API. The Spritzer dataset is widely used in computational social science because it is drawn at the platform level rather than via keyword or user-based queries, making it broadly representative of public Twitter activity with respect to volume, language mix, and topical distribution. The data were obtained from the Archive Team Twitter Stream Grab of the Internet Archive. Our window covers three months in 2022 and one month in 2023, chosen because they overlap with our trending topic collection. To ensure comparability with the trending topic data, we keep only German-speaking tweets in the 1\% sample. This yields \nspritzertweets tweets for the four months.  

\subsection{Data completeness}
We collected tweets using the Twitter Search API v1, which has been shown to provide only a certain sample of all tweets (see \citep{Pournaki2021} for a comparison between Search and Stream API which shows that the Search API contains 80\% of the tweets collected via the Stream API). Also, we collected each trending keyword for maximally 24h, thereby potentially leaving out tweets at the very onset of a trend. Finally, the collection was inactive between 2022-07-30 and 2022-09-03, which results in a total uptime of 96\%. 
\clearpage
\section{Human annotation procedure}
\label{sec:annotation_procedure}
\begin{figure}[h]
    \includegraphics[width=\linewidth]{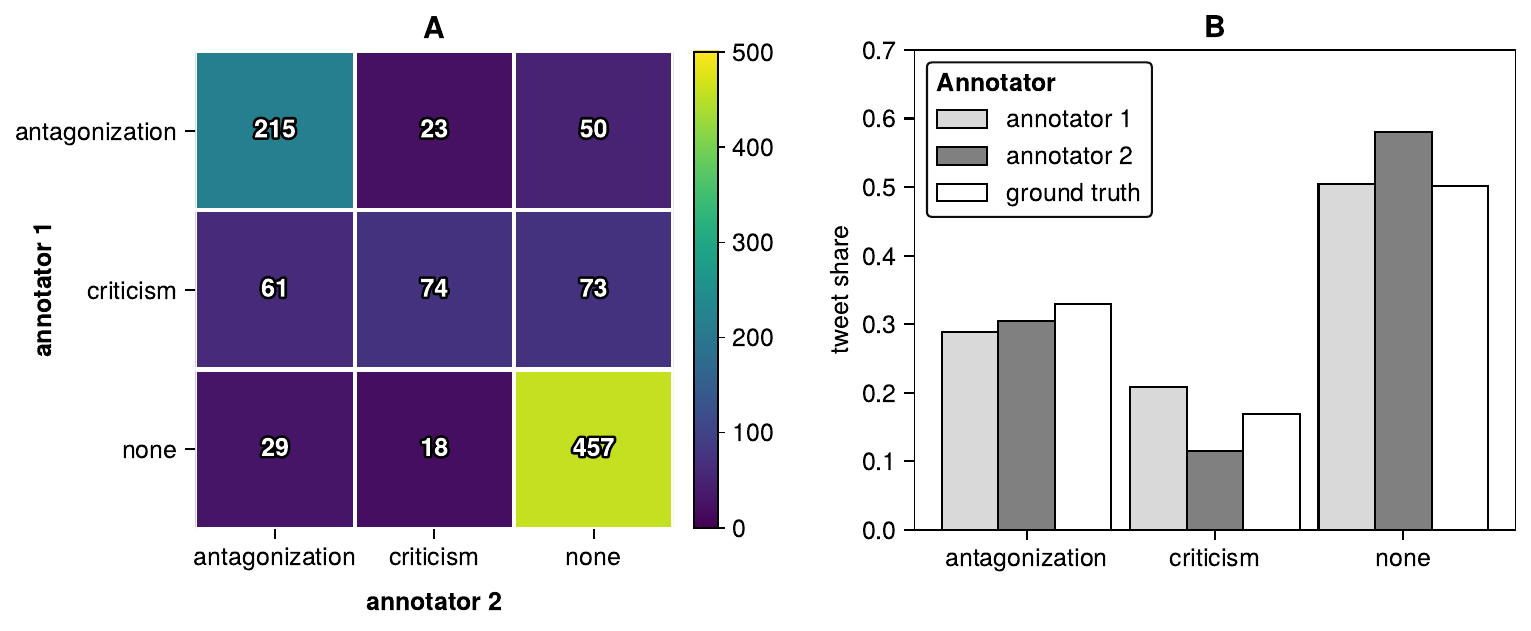}
    \caption{\textbf{Inter-annotator agreement and reconciliation.} (A) Confusion matrix of the labels assigned independently by the two annotators to the 1000 annotated tweets (raw agreement 74.6\%, Cohen's $\kappa = 0.57$). (B) Share of each category in the labels of each annotator and in the reconciled ground truth.}    
    \label{fig:human_confusion_matrix}
\end{figure}
In the initial annotation phase, the two authors, German native speakers with domain knowledge in German politics, annotated 1000 randomly selected tweets for antagonization/criticism/none, as well as the potential targets. This yielded an agreement score of $\kappa=0.57$. 

The confusion matrix in Fig.\ref{fig:human_confusion_matrix} (panel A) reveals that the identification of criticism is particularly difficult. We identify two effects. On the one hand, looking at those cases where annotator 1 assigned ``none'' and annotator 2 ``criticism'', the criticism is often times implicit and contextual, which led annotators to disagree. On the other hand, looking at those cases where annotator 1 assigned ``criticism'' and annotator 2 ``antagonization'', the disagreement most often was about the limits between criticism and ridicule. Phrases such as ``panicmongering'' in the case of Covid, or sarcastic descriptions are at the boundary between antagonization and criticism. These edge cases highlight the difficulty of the annotation task. 

In a final step, the disagreements were reconciled in mutual discussion, yielding the ``ground truth'' dataset with the distribution of antagonization/criticism/none shown in Fig.~\ref{fig:human_confusion_matrix} (panel B).

\clearpage

\section{Prompt}
\label{sec:prompt}
{
    \small
    \ttfamily
You are an expert in discourse analysis with a focus on political and social media rhetoric. Your task is to classify the author’s stance (antagonization/criticism/none) in the following text with maximum fidelity to human annotation standards. Do not guess. Base your judgment strictly on the author’s own words, not on attributed opinions or reported speech.\\

- Antagonization means portraying a person, group, or entity as an enemy, threat, or hostile “other,” OR using contemptuous, demeaning personal attacks (e.g., insults, ridicule, belittlement, dehumanizing metaphors) that delegitimize the target. It often involves emotional language, us-vs-them framing, or discrediting the person rather than their ideas.\\
- Criticism means expressing objections, disagreement, or negative evaluation without enemy/hostile framing and without personal disparagement; it focuses on actions, decisions, or ideas.
- Only classify a passage as antagonization or criticism if the author of the text is taking that stance. If the author merely reports, quotes, or attributes antagonization/criticism to someone else, set "stance": "none".\\

\# Analysis Steps\\
1. Stance Detection\\
Determine if the author expresses:\\
- "antagonization" → if the author portrays the target as an enemy/hostile other, uses demeaning personal attacks, ridicule, or delegitimizing language.\\
- "criticism" → if the author expresses objection or disapproval without enemy framing or personal disparagement.\\
- "none" → if neither is present or the statement is merely attributed/quoted.\\

2. Target Identification (if antagonization or criticism)\\
- Identify the specific target (individual, group, or entity) that the author directly antagonizes or criticizes.\\
- Targets can be:\\
\null\quad - Individuals\\
\null\quad - Groups\\
\null\quad - Institutions/Entities\\
\null\quad - Ideas/Policies when the critique is idea-focused\\
- Do NOT include entities that are only:\\
\null\quad - mentioned as objects of someone else's criticism/antagonization\\
\null\quad - referenced neutrally\\
- Quote directly from the text.\\
- DO NOT translate the target.\\
- If there is more than one target, delimit them by "|".\\

\# Examples\\
Example 1 (antagonization)\\
Text: "Die Medien lügen ständig und versuchen, das Volk zu manipulieren."\\
\{"stance": "antagonization","target": "Medien"\}\\

Example 2 (antagonization)\\
Text: "Jetzt alle! \#SuspendXYZ \#fckABC"\\
\{"stance":"antagonization","target":"XYZ|ABC"\}\\
Example 3 (criticism)\\
Text: "Die Regierung hat in der Klimapolitik Fehler gemacht und sollte ihre Strategie überdenken."\\
\{"stance": "criticism","target": "Regierung"\}\\

Example 4 (criticism)\\
Text: "Lasst die Patienten nicht länger leiden, \texttt{@}spdbt \texttt{@}Die\_Gruenen \texttt{@}fdp"\\
\{"stance":"criticism","target":"\texttt{@}spdbt|\texttt{@}Die\_Gruenen|\texttt{@}fdp"\}\\

Example 5 (none)\\
Text: "Morgen findet eine Konferenz zur Energiewende statt."\\
\{"stance": "none","target": null\}\\

Example 6 (none)\\
Text: "Einige Kritiker behaupten, die Regierung sei korrupt."\\
\{"stance": "none","target": null\}\\

\# Output Format (strict JSON)\\
\{"stance": "antagonization" | "criticism" | "none",\\
 "target": "…" | null\}\\

Additional Consistency Rule:\\
- If "stance" is "antagonization" or "criticism", "target" must not be null.
}
\clearpage

\section{Model selection}
\label{sec:appendix:model-selection}
\begin{table}[h]
\caption{\textbf{Model comparison.} Performance of the evaluated open-weight models on the 1000 annotated tweets, sorted by macro F1 score.}
\label{tab:f1scores}
\begin{tabular}{lrrrrr}
\toprule
 & antagonization (F1) & criticism (F1) & none (F1) & macro avg (F1) & weighted avg (F1) \\
model &  &  &  &  &  \\
\midrule
qwen3:32b & 0.76 & 0.50 & 0.83 & 0.70 & 0.75 \\
llama3.3:70b & 0.65 & 0.54 & 0.82 & 0.67 & 0.72 \\
qwen3:14b & 0.70 & 0.51 & 0.81 & 0.67 & 0.72 \\
phi4:14b & 0.69 & 0.52 & 0.81 & 0.67 & 0.72 \\
qwen2.5:14b & 0.72 & 0.43 & 0.79 & 0.65 & 0.71 \\
llama3.1:70b & 0.65 & 0.53 & 0.80 & 0.66 & 0.71 \\
qwen2.5:72b & 0.71 & 0.50 & 0.78 & 0.66 & 0.71 \\
qwen3:8b & 0.71 & 0.46 & 0.77 & 0.64 & 0.70 \\
qwen2.5:32b & 0.65 & 0.51 & 0.81 & 0.66 & 0.70 \\
qwen3:30b & 0.67 & 0.42 & 0.80 & 0.63 & 0.70 \\
qwen2.5:7b & 0.64 & 0.48 & 0.80 & 0.64 & 0.69 \\
mistral-small3.2:24b & 0.61 & 0.48 & 0.78 & 0.62 & 0.67 \\
ministral-3:8b & 0.71 & 0.42 & 0.71 & 0.61 & 0.66 \\
gemma3:27b & 0.69 & 0.49 & 0.66 & 0.61 & 0.64 \\
ministral-3:14b & 0.67 & 0.46 & 0.66 & 0.60 & 0.63 \\
phi4-mini:3.8b & 0.59 & 0.44 & 0.69 & 0.57 & 0.61 \\
qwen3:4b & 0.68 & 0.39 & 0.63 & 0.56 & 0.60 \\
qwen2.5:3b & 0.64 & 0.11 & 0.71 & 0.48 & 0.58 \\
ministral-3:3b & 0.62 & 0.06 & 0.67 & 0.45 & 0.55 \\
gemma3:12b & 0.71 & 0.43 & 0.49 & 0.54 & 0.55 \\
qwen3:0.6b & 0.42 & 0.31 & 0.67 & 0.47 & 0.53 \\
qwen3:1.7b & 0.63 & 0.31 & 0.45 & 0.46 & 0.49 \\
llama3.1:8b & 0.59 & 0.17 & 0.53 & 0.43 & 0.49 \\
llama3.2:1b & 0.51 & 0.01 & 0.62 & 0.38 & 0.48 \\
mixtral:8x7b & 0.38 & 0.40 & 0.57 & 0.45 & 0.48 \\
qwen2.5:1.5b & 0.55 & 0.02 & 0.47 & 0.35 & 0.42 \\
mistral-nemo:12b & 0.41 & 0.20 & 0.48 & 0.37 & 0.41 \\
llama3.2:3b & 0.55 & 0.01 & 0.39 & 0.32 & 0.38 \\
qwen2.5:0.5b & 0.11 & 0.01 & 0.66 & 0.26 & 0.37 \\
mistral:7b & 0.20 & 0.30 & 0.44 & 0.31 & 0.34 \\
gemma3:270m & 0.49 & 0.00 & 0.29 & 0.26 & 0.31 \\
gemma3:4b & 0.57 & 0.20 & 0.15 & 0.31 & 0.30 \\
gemma3:1b & 0.51 & 0.16 & 0.00 & 0.22 & 0.20 \\
gpt-oss:20b & 0.01 & 0.00 & 0.37 & 0.13 & 0.19 \\
\bottomrule
\end{tabular}
\end{table}
\begin{table}[h!]
\caption{\textbf{Performance of qwen3:32b.} Precision, recall and F1 score per category, evaluated against human annotation ($n = 1000$), together with macro and weighted averages.}
\label{tab:qwenf1}
\begin{tabular}{lrrrr}
\toprule
 & precision & recall & f1-score & support \\
\midrule
antagonization & 0.74 & 0.77 & 0.76 & 330 \\
criticism & 0.59 & 0.44 & 0.50 & 169 \\
none & 0.81 & 0.86 & 0.83 & 501 \\
accuracy & 0.76 & 0.76 & 0.76 & 0.76 \\
macro avg & 0.71 & 0.69 & 0.70 & 1000 \\
weighted avg & 0.75 & 0.76 & 0.75 & 1000 \\
\bottomrule
\end{tabular}
\end{table}
\begin{figure}[h!]
    \includegraphics[width=.6\linewidth]{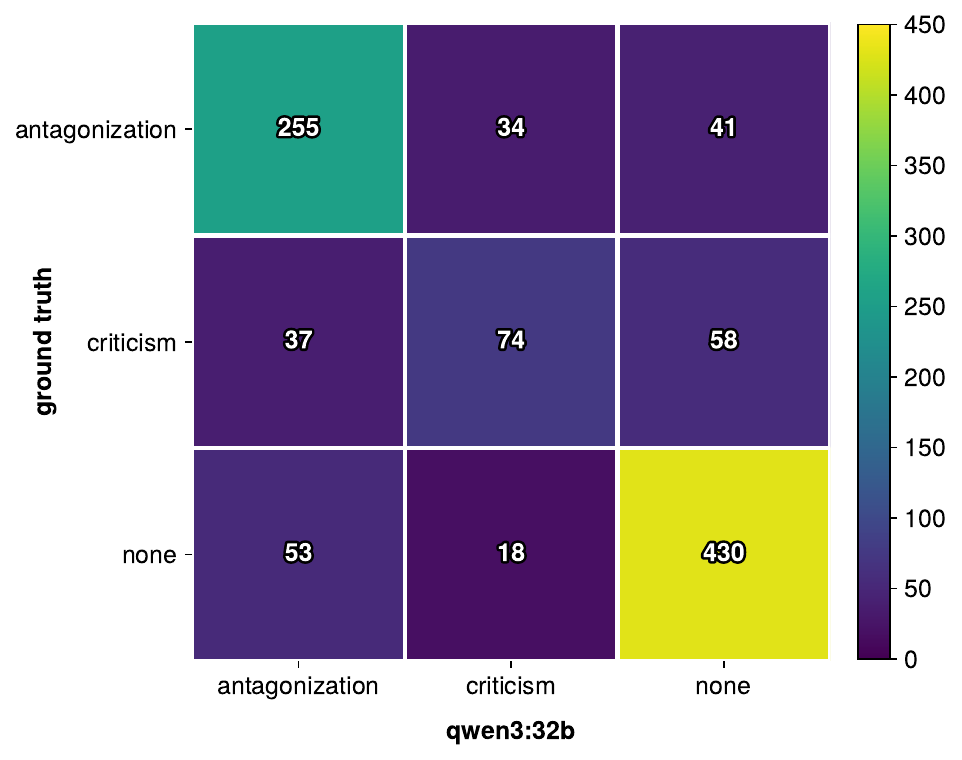}
    \caption{\textbf{Confusion matrix of the human-annotated ground truth and qwen3:32b.} Rows give the human annotation, columns the label assigned by the model ($n = 1000$), such that diagonal cells correspond to agreement.}
    \label{fig:confusion_human_qwen}
\end{figure}

\begin{table}[htbp]
\caption{Confusion matrices of human annotation and classifier labels for the annotated tweets of each political camp. Rows give the human annotation, columns the label assigned by the classifier, so that diagonal cells correspond to agreement.}
\label{tab:confusion_camps}
\begin{tabular}{llrrr}
\toprule
 & & \multicolumn{3}{c}{qwen3:32b} \\
\cmidrule(lr){3-5}
Political leaning & Human annotation & Antagonization & Criticism & None \\
\midrule
Left-leaning (n~=~377) & Antagonization & 70 & 13 & 15 \\
 & Criticism & 14 & 25 & 24 \\
 & None & 16 & 10 & 190 \\
\midrule
Right-leaning (n~=~344) & Antagonization & 129 & 14 & 23 \\
 & Criticism & 10 & 25 & 13 \\
 & None & 18 & 7 & 105 \\
\midrule
N.A. (n~=~279) & Antagonization & 56 & 7 & 3 \\
 & Criticism & 13 & 24 & 21 \\
 & None & 19 & 1 & 135 \\
\bottomrule
\end{tabular}
\end{table}
\clearpage
\section{Validation of target extraction}
\label{sec:appendix:target-evaluation}
\begin{table}[htbp]
\caption{Manual verification of the extracted targets for the 400 annotated tweets to which both the human annotation and the classifier assigned antagonization or criticism. Targets are considered correct if all targets addressed in the tweet were extracted and no additional targets were added.}
\label{tab:target_evaluation}
\begin{tabular}{lr}
\toprule
\textbf{Evaluation} & \textbf{n (\%)} \\
\midrule
Correct              & 337 (84\%) \\
Target(s) false  &  47 (12\%) \\
Target missing       &   7 (2\%) \\
Target added         &   6 (2\%) \\
One target false     &   2 ($<$1\%) \\
Unresolved pronouns  &   1 ($<$1\%) \\
\midrule
Total                & 400 (100\%) \\
\bottomrule
\end{tabular}
\end{table}
To evaluate the target extraction, we manually verified the extracted targets on a subset of our ground truth. We selected all tweets among the 1000 annotated tweets for which both the human annotation and the classifier assigned either antagonization or criticism, since only these tweets contain targets to compare ($n=400$). For each of these tweets, we checked whether the targets extracted by the model matched the targets addressed in the tweet, and categorized the outcome as shown in Table~\ref{tab:target_evaluation}. The extracted targets were correct in 84\% of cases. Since this evaluation is restricted to tweets on which annotation and classifier agree, it measures the quality of the target extraction independently of classification errors.

\section{Target merging}
\label{sec:target_merges}
For each antagonizing or criticizing tweet, the model returns one or more targets. The same target can be referred to in different ways across the corpus (e.g., the politician Armin Laschet appears both as ``Armin Laschet'' and as ``Laschet''), which would split its occurrences across several strings. To avoid this, we merged target variants manually. After preprocessing the target strings (removing trailing spaces and hashtag symbols, and replacing Twitter handles with the corresponding account names), we examined the 50 most frequent targets for each political camp. For each of these targets, we searched the corpus for other strings referring to the same entity and merged them. This yielded the following merged targets:

{\small
\textbf{afd}: afd, noafd, fckafd, afd-fraktion im deutschen bundestag \\
\textbf{armin laschet}: armin laschet, laschet, herr laschet, arminlaschet, herrn laschet, armin laschet (cdu) \\
\textbf{cdu}: cdu, cdu deutschlands, cdu-twitter \\
\textbf{cdu/csu}: cdu/csu, cdu/csu-fraktion, cducsu, cdu csu, cdu \& csu, union \\
\textbf{querdenker}: querdenker, querdenkern, querdenker*innen, querdenker:innen, querdenker-sekte \\
\textbf{vladimir putin}: putin, vladimir putin, russischen präsidenten putin, herr putin \\
\textbf{nazis}: nazis, nazi, nzs \\
\textbf{neonazis}: neonazis, neo-nazis \\
\textbf{russland}: rußland, russland \\
\textbf{sahra wagenknecht}: wagenknecht, sahra wagenknecht, frau wagenknecht, sarah wagenknecht, fr. wagenknecht, swagenknecht, sahra\_wagenknecht, s. wagenknecht \\
\textbf{friedrich merz}: merz, friedrich merz, herr merz, \_friedrichmerz, herr \_friedrichmerz, herr friedrich merz, herrn merz, friedrichmerz \\
\textbf{polizei}: polizei, deutsche polizei, bundespolizei, polizeiapparat \\
\textbf{christian lindner}: lindner, christian lindner, herr lindner, c\_lindner, finanzminister lindner, \#fdp-chef c\_lindner, herr c\_lindner, c. lindner, herrn lindner, fdp-chef lindner, fdp lindner, fdp-chef c\_lindner \\
\textbf{spd}: spd, spd-fraktion im bundestag, spd parteivorstand, spdbt, spdde \\
\textbf{hans-georg maaßen}: maaßen, hans-georg maaßen, herr maaßen, hans georg maaßen, hg maaßen, h.g. maaßen, herrn maaßen, h.-g. maaßen \\
\textbf{olaf scholz}: scholz, olaf scholz, bundeskanzler olaf scholz, herr scholz, kanzler scholz, bundeskanzler scholz, olafscholz, bundeskanzler \#scholz, herrn scholz, bundeskanzler olafscholz, bundesfinanzminister olaf scholz, olaf\_scholz \\
\textbf{die grünen}: grünen, bündnis 90/die grünen, die grünen, diegrünen, gruenen, grüne im bundestag, grüne partei, gruene, grüne, diegruenen, die\_gruenen, gruenebw, gruenebundestag \\
\textbf{russen}: russen, russe \\
\textbf{annalena baerbock}: baerbock, außenministerin annalena baerbock, annalena baerbock, frau baerbock, deutsche außenministerin abaerbock, annalenabaerbock, kanzlerinbaerbock, anna-lena baerbock, außenministerin baerbock, baerbocks, aussenministerin baerbock, abaerbock, frau abaerbock, a. baerbock, fr. baerbock, kanzlerin baerbock \\
\textbf{karl lauterbach}: lauterbach, karl lauterbach, prof. karl lauterbach, herr lauterbach, gesundheitsminister lauterbach, bundesgesundheitsminister lauterbach, karl\_lauterbach, herrn lauterbach, herr karl\_lauterbach, karllauterbach, hr lauterbach \\
\textbf{robert habeck}: habeck, robert habeck, wirtschaftsminister habeck, herr habeck, herrn habeck, minister habeck \\
\textbf{medien}: medien, deutsche medien, staatsmedien, leitmedien, mainstream-medien, massenmedien, deutschen medien, mainstreammedien, öffentlichen-rechtlichen medien, medien(-vertreter), mainstream medien, qualitätsmedien, medienwelt, deutsche qualitätsmedien \\
\textbf{politiker}: politiker, deutsche politiker, politiker*innen, politikern, politiker:innen, politiker des parteienkartells, politikerinnen, unsere politiker, politikerïnnen, politikerin, spitzenpolitiker, eine politikerin, berufspolitiker, deutschen politikern, deutsche politikerklasse, ein politiker, aktive politiker, deutscher politiker, politikerinnen und politiker, führenden politikern, deutschen politiker \\
\textbf{ungeimpfte}: ungeimpfte, ungeimpften, ungeimpfte leute, ungeimpfte menschen, ungeimpfte personen, ungeimpfte (ungeimpfte), ungeimpfter, ungeimpfte kollegen \\
\textbf{bundesregierung}: bundesregierung, deutsche bundesregierung, bureg, regierung, deutsche regierung, unsere regierung, diese regierung, regierungen \\
\textbf{angela merkel}: merkel, angela merkel, frau merkel, kanzlerin merkel, frau dr. merkel \\
\textbf{ampel-regierung}: ampel, ampel-regierung, ampelregierung, ampelkoalition, ampel-koalition \\
\textbf{tagesschau}: tagesschau, ard-tagesschau, tagesschau.de, tagesschau24, medien (tagesschau), ard tagesschau, tagesschau ard \\
\textbf{jens spahn}: spahn, jens spahn, gesundheitsminister spahn, herr spahn, jensspahn, herrn spahn, herr jensspahn, minister spahn, gesundheitsminister jensspahn, bundesgesundheitsminister spahn, spahns, j. spahn, hr. spahn \\
\textbf{migrants/refugees}: flüchtlinge, flüchtlingen, afghanische flüchtlinge, vermeintliche flüchtlinge, flüchtlinge aus der 3. welt, flüchtlinge aus afghanistan, flüchtling, flüchtling aus eritrea, geflüchtete, 30-jähriger syrischer flüchtling, flüchtlingswelle, flüchtlinge/frauen und kinder aus afghanistan, flüchtlinge des ukrainekrieg, migranten, junge männer mit migrationshintergrund, migrantengruppen, menschen mit migrationshintergrund, mädchen mit migrationshintergrund, illegale migranten, migrant, muslimische migranten, migranten mit migrationshintergrund, migrantische milieu, franzosen mit migrationshintergrund, personen mit migrationshintergrund, migranten (implizit durch kontext), jungen männern mit migrationshintergrundeinwanderer, algerische einwanderin, muslimische einwanderer, illegale einwanderer, einwandernder sozialer bodensatz, unzivilisierte einwanderer, muslimischen einwanderern, illegale und häufig aggressive einwanderer aus anderen kulturen, kriminelle einwanderer
}
\clearpage
\section{Targets of criticism}
\begin{table}[h]
\caption{\textbf{Top targets of criticism by leaning.}}
\label{tab:targets_criticism}
\begin{tabular}{llrlr}
\toprule
 & \multicolumn{2}{c}{LEFT} & \multicolumn{2}{c}{RIGHT} \\
 & target & count & target & count \\
\midrule
1 & Armin Laschet & 90623 & Government & 52892 \\
2 & CDU & 64964 & Green Party & 44571 \\
3 & FDP & 42058 & Annalena Baerbock & 44250 \\
4 & CDU/CSU & 36759 & Robert Habeck & 35383 \\
5 & SPD & 35522 & Karl Lauterbach & 33191 \\
6 & Government & 34637 & Olaf Scholz & 25037 \\
7 & Olaf Scholz & 32912 & Traffic light coalition & 22858 \\
8 & Green Party & 32165 & Media & 20083 \\
9 & Police & 21372 & Politics & 18167 \\
10 & Politics & 21194 & Deutschland & 15182 \\
11 & Jens Spahn & 20286 & Politicians & 14711 \\
12 & Christian Lindner & 20278 & SPD & 11919 \\
13 & CSU & 16334 & Vaccination mendate & 11693 \\
14 & Friedrich Merz & 15630 & FDP & 10160 \\
15 & Annalena Baerbock & 12861 & CDU & 10137 \\
16 & Politicians & 12532 & Jens Spahn & 8989 \\
17 & Traffic light coalition & 12525 & Angela Merkel & 8696 \\
18 & UEFA & 11078 & Vaccine & 8292 \\
19 & Germany & 10834 & ZDF & 8171 \\
20 & Media & 10494 & State & 7585 \\
\bottomrule
\end{tabular}
\end{table}
\clearpage

\section{Statistical uncertainty quantification via prediction-powered inference (PPI)}
\label{sec:appendix:ppi}

Our corpus consists of $N = \ntweets$ tweets, each assigned a label in $\{\mathrm{antagonization}, \mathrm{criticism}, \mathrm{none}\}$ by an LLM-based classifier $f$. The quantities of interest are proportions of the form $p_c = \mathbb{P}(Y = c)$, where $Y$ denotes the human-judged label, as well as proportions conditional on auxiliary variables (political leaning and time). Because the classifier is imperfect, the naive plug-in estimate $\tilde{p}_c = \frac{1}{N}\sum_{i=1}^{N} \mathbf{1}\{f(X_i) = c\}$ is biased: it reflects the distribution of LLM-predicted labels, not human-judged ones. 

To correct for this, we use a human-annotated validation set of $n = 1000$ tweets, sampled uniformly at random from the corpus, for which both the true label $Y_i$ and the LLM's prediction $f(X_i)$ are available. Prediction-powered inference (PPI; \citet{Angelopoulos2023}) combines this labeled validation set with the large unlabeled corpus to produce confidence intervals for proportions that are valid despite classifier error. For a given proportion, we define the \emph{rectifier} as the classifier's mean signed error on the validation set,
\begin{equation}
\label{eq:ppi_rect}
\hat{R}_c = \frac{1}{n}\sum_{i \in \mathcal{V}} \Big( \mathbf{1}\{f(X_i) = c\} - \mathbf{1}\{Y_i = c\} \Big),
\end{equation}
estimated on the validation set $\mathcal{V}$. The PPI point estimate subtracts this systematic error from the naive proportion computed on the full corpus,
\begin{equation}
\label{eq:ppi_prop}
\hat{p}_c^{\,\mathrm{PPI}} = \tilde{p}_c - \hat{R}_c,
\end{equation}
and the associated confidence interval combines the sampling variance of both terms,
\begin{equation}
\label{eq:ppi_conf}
\widehat{\mathrm{Var}}\big(\hat{p}_c^{\,\mathrm{PPI}}\big) = \frac{\widehat{\mathrm{Var}}_{\mathcal{U}}\big(\mathbf{1}\{f(X) = c\}\big)}{N} + \frac{\widehat{\mathrm{Var}}_{\mathcal{V}}\big(\mathbf{1}\{f(X) = c\} - \mathbf{1}\{Y = c\}\big)}{n},
\end{equation}
where $\mathcal{U}$ denotes the unlabeled corpus. With $N \gg n$, the first term is negligible and the interval width is governed almost entirely by the validation set: uncertainty about the classifier's error pattern, not about sampling from the corpus, dominates.
\paragraph{Marginal class proportions}
Each marginal proportion is the mean of a one-vs-rest indicator, so we apply PPI for the mean three times, with $Y_i^{(c)} = \mathbf{1}\{Y_i = c\}$ and $f^{(c)}(X_i) = \mathbf{1}\{f(X_i) = c\}$ for $c \in \{\mathrm{antagonization}, \mathrm{criticism}, \mathrm{none}\}$. The three point estimates sum to one by construction, but the three intervals are computed independently and are not jointly calibrated.

Comparing prediction-powered estimates to the classifier's shares quantifies the classifier's bias. Across all tweets, the classifier understates criticism by 4.3 percentage points and overstates ``none'' and antagonization by 2.8 and 1.5 percentage points, respectively. This reflects that tweets annotated as criticism are frequently classified as ``none'' or as antagonization (Table~\ref{tab:confusion_camps}). Within the two political camps, the corrections do not exceed 4 percentage points, whereas for tweets from unassigned accounts, the classifier overstates antagonization by 7.9 and understates criticism by 9.3 percentage points. All corrections are reported in Table~\ref{tab:ppi_corrections}.

\paragraph{Proportions within political camps}
For proportions within political camps, $\mathbb{P}(\mathrm{class} = c \mid \mathrm{camp} = l)$ with $l \in \{\mathrm{left}, \mathrm{right}, \mathrm{not~assigned}\}$, we treat the assignment of accounts to political camps as error-free. Under this assumption, the problem decomposes by camp: we partition both the corpus and the annotated tweets by camp and apply prediction-powered inference within each camp, so that the correction for each camp is estimated only from the annotated tweets of that camp. This is possible because the annotated tweets were sampled at random within each camp, and it avoids assuming that the classifier's errors are the same across political camps.

\paragraph{Temporal analysis}
For the monthly time series of $\hat{p}_{\mathrm{antagonization},t}$, we partition the corpus into monthly bins and apply PPI within each bin, reusing the full validation set to estimate the rectifier in every bin. This embeds a stationarity assumption: the classifier's error distribution is constant over the study period. 

\paragraph{Relationship to bootstrapping}
A natural alternative for uncertainty quantification would be a nonparametric bootstrap over the $N = \ntweets$ classified tweets. We do not report bootstrap intervals for a quantitative reason. The bootstrap quantifies sampling variability only: it treats the classifier's output as ground truth and asks how the estimated proportions would vary across resamples of the corpus. For a proportion at this corpus size, the bootstrap standard error coincides with the closed-form multinomial value $\sqrt{\tilde{p}(1-\tilde{p})/N} \approx 1.1 \times 10^{-4}$ (for $\tilde{p} \approx 0.34$), yielding 95\% intervals of roughly $\pm 0.0002$, approximately two orders of magnitude narrower than the PPI intervals of about $\pm 0.025$ reported above.

This discrepancy is important: total uncertainty in our estimates is dominated by classifier measurement error, which the bootstrap does not capture. Moreover, the bootstrap's contribution is not omitted from the PPI analysis, as the corpus-sampling variance appears as the first (negligible) term in the PPI variance decomposition in Eq.~\eqref{eq:ppi_conf}. Reporting bootstrap intervals alongside PPI intervals would therefore present a strict subset of the uncertainty already accounted for, and presenting them alone would understate uncertainty by orders of magnitude.

\paragraph{Assumptions and limitations}

Our inferential framework rests on the following assumptions, stated here for transparency. First, the validation set is an i.i.d. sample from the same population as the corpus, with the reconciled human annotations treated as ground truth. Second, the political leaning classifier is treated as error-free; errors in that classifier would propagate to joint and conditional estimates and are not corrected. Third, temporal analyses assume a stationary classifier error distribution. 

\paragraph{Implementation}
We implement prediction-powered inference using the Python package \texttt{ppi\_py} \citep{Angelopoulos2023}. Throughout, we use classical prediction-powered inference with the power-tuning parameter fixed at $\lambda = 1$.

\begin{table}[t]
\caption{Classifier shares and prediction-powered estimates of each category, for all tweets and within each political camp. Within camps, corrections are estimated separately, using only the annotated tweets of that camp ($n_{\text{val}}$). The correction is the prediction-powered estimate minus the classifier share, in percentage points, so that positive values indicate that the classifier under-predicts a category.}
\label{tab:ppi_corrections}
\begin{tabular}{llrrcr}
\toprule
Political leaning & Category & $n_{\text{val}}$ & Classifier (\%) & PPI (\%) [95\% CI] & Correction (pp) \\
\midrule
All tweets & Antagonization & 1000 & 34.2 & 32.7 [30.2, 35.3] & -1.5 \\
 & Criticism &  & 16.8 & 21.1 [18.7, 23.4] & +4.3 \\
 & None &  & 49.0 & 46.2 [43.6, 48.7] & -2.8 \\
\midrule
Left-leaning & Antagonization & 377 & 28.9 & 28.4 [24.4, 32.3] & -0.5 \\
 & Criticism &  & 19.8 & 23.8 [19.8, 27.9] & +4.0 \\
 & None &  & 51.3 & 47.8 [43.6, 52.0] & -3.4 \\
\midrule
Right-leaning & Antagonization & 344 & 46.2 & 48.8 [44.2, 53.4] & +2.6 \\
 & Criticism &  & 18.3 & 18.9 [15.1, 22.6] & +0.6 \\
 & None &  & 35.5 & 32.3 [27.9, 36.8] & -3.2 \\
\midrule
N.A. & Antagonization & 279 & 26.8 & 18.9 [14.4, 23.3] & -7.9 \\
 & Criticism &  & 10.8 & 20.1 [15.7, 24.5] & +9.3 \\
 & None &  & 62.5 & 61.0 [56.4, 65.7] & -1.4 \\
\bottomrule
\end{tabular}
\end{table}

\clearpage

\section{Number of tweets per topic}
\begin{figure}[h]
    \includegraphics[width=\linewidth]{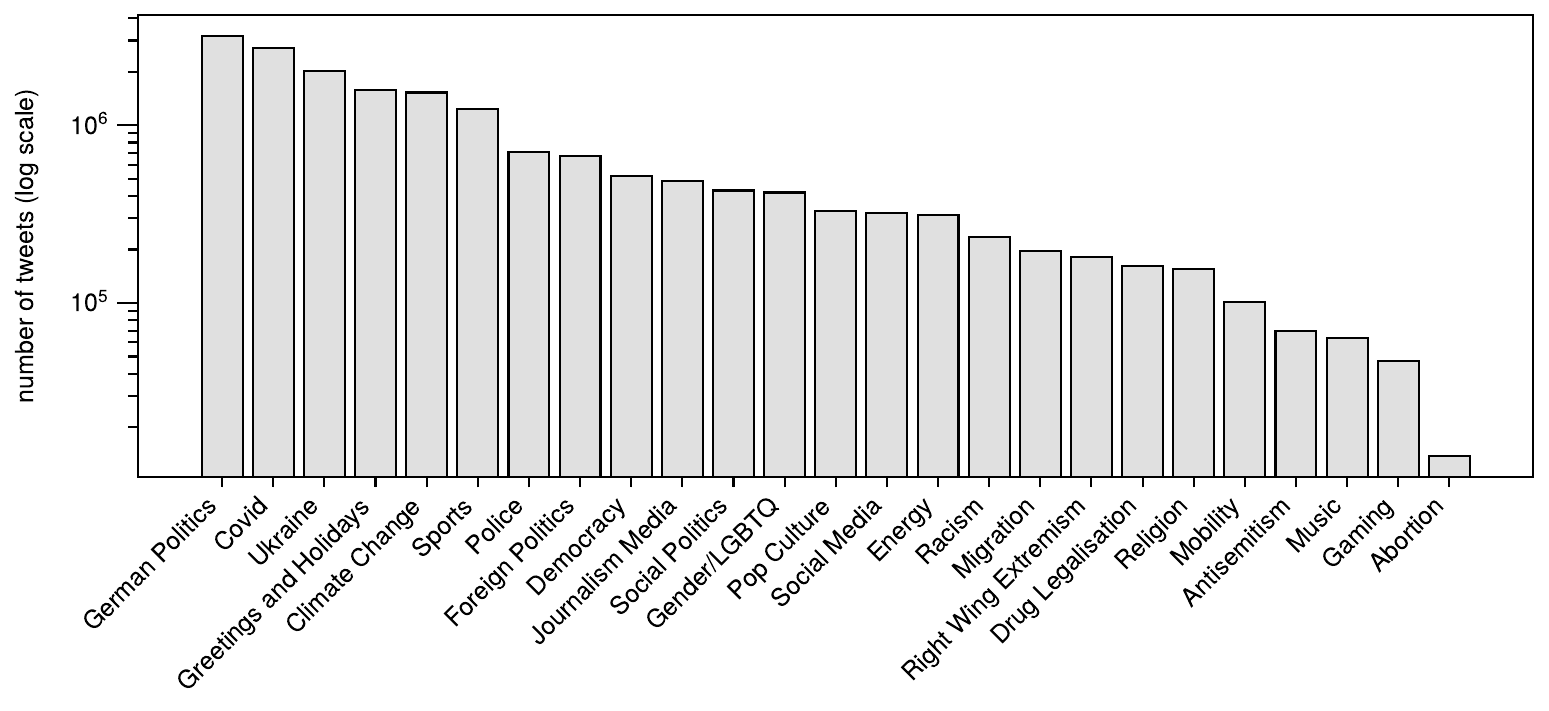}
    \caption{\textbf{Number of tweets per topic.}}
    \label{fig:n_retweets_per_topic}
\end{figure}

\begin{figure}[h]
    \includegraphics[width=\linewidth]{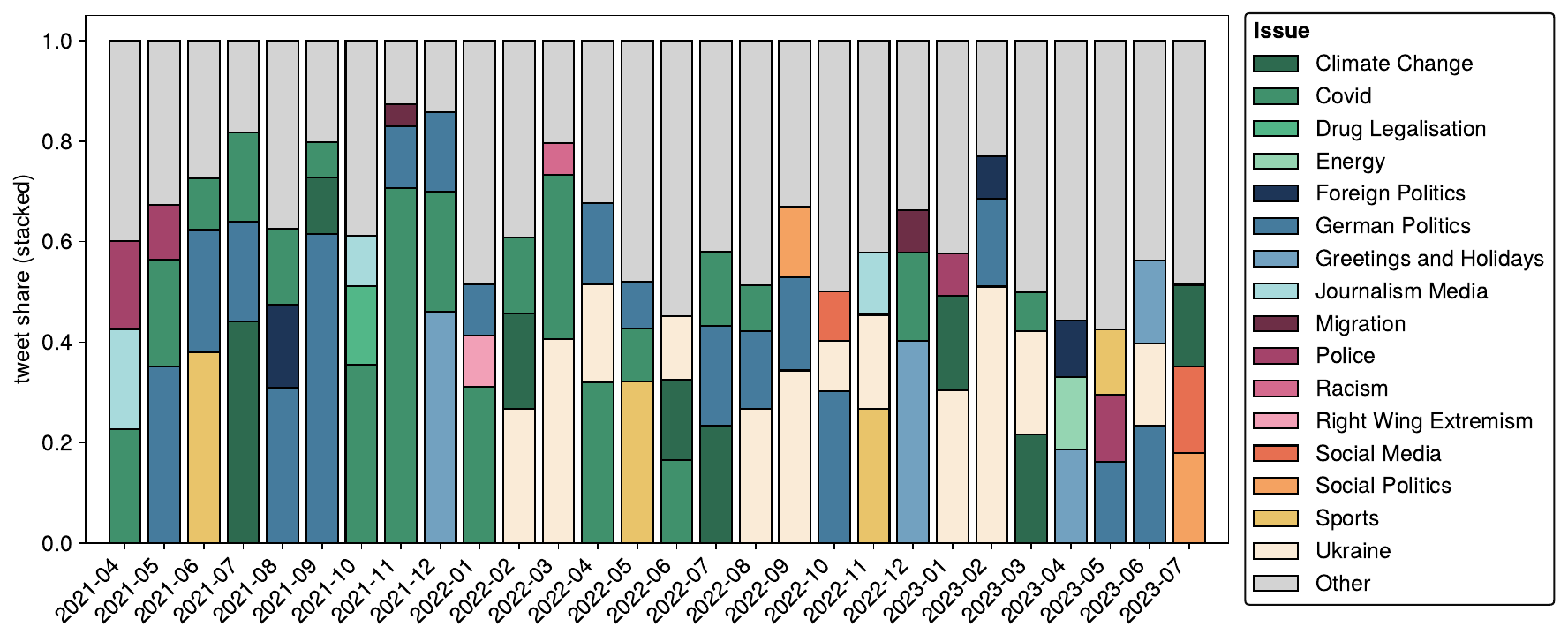}
    \caption{\textbf{Topic-wise tweet share over time.}}
    \label{fig:topics_over_time}
\end{figure}
\clearpage

\clearpage
\section{Codebook for identifying antagonization and criticism in tweets}

\begin{enumerate}
\def\labelenumi{\arabic{enumi})}
\tightlist
\item
  Purpose and Scope
\end{enumerate}

\begin{itemize}
\tightlist
\item
  Goal: Consistently label tweets for antagonization, criticism, or
  none, and identify the direct target(s) when applicable.
\item
  Domain: Short social media texts (tweets), including replies, quote
  tweets, and tweets with hashtags, mentions, emojis, or links.
\item
  Language: Annotate in the language of the tweet. Do not translate
  quoted targets.
\end{itemize}

\begin{enumerate}
\def\labelenumi{\arabic{enumi})}
\setcounter{enumi}{1}
\tightlist
\item
  Unit of Analysis
\end{enumerate}

\begin{itemize}
\tightlist
\item
  One tweet is the annotation unit.
\item
  For quote tweets, annotate only the author's commentary as their
  stance. Treat the quoted content as attributed speech unless the
  author explicitly endorses it.
\end{itemize}

\begin{enumerate}
\def\labelenumi{\arabic{enumi})}
\setcounter{enumi}{2}
\tightlist
\item
  Labels and Definitions
\end{enumerate}

\begin{itemize}
\tightlist
\item
  antagonization:

  \begin{itemize}
  \tightlist
  \item
    Definition: The author portrays a person, group, or entity as an
    enemy, threat, or hostile ``other,'' and/or uses contemptuous,
    demeaning personal attacks (insults, ridicule, belittlement,
    dehumanizing metaphors) that delegitimize the target.
  \item
    Cues: Personal insults (``idiot,'' ``Lügner,'' ``korrupt(e)
    Schweine''), ridicule, name-calling, dehumanization (``Ratten,''
    ``Ungeziefer''), enemy/threat framing (``Feinde,'' ``Invasoren,''
    ``Bedrohung''), dismissing legitimacy (``gehört weg'').
  \end{itemize}
\item
  criticism:

  \begin{itemize}
  \tightlist
  \item
    Definition: The author expresses objections, disagreement, or
    negative evaluation without enemy/hostile framing and without
    personal disparagement. In contrast to antagonization, the
    (political) opponent is acknowledged as legitimate. Focus is on
    actions, decisions, policies, or ideas rather than personal traits.
  \item
    Cues: Specific issues (``hat Fehler gemacht,'' ``schlechte
    Politik,'' ``Strategie ist unklar''), calls for improvement or
    change, reasoned disagreement.
  \end{itemize}
\item
  none:

  \begin{itemize}
  \tightlist
  \item
    Definition: No authorial antagonization or criticism. This includes
    neutral or positive content, unrelated content, and cases where
    antagonization/criticism is merely reported, quoted, or attributed
    to others without clear endorsement by the author.
  \end{itemize}
\end{itemize}

\begin{enumerate}
\def\labelenumi{\arabic{enumi})}
\setcounter{enumi}{3}
\tightlist
\item
  Label Precedence (Tie-Breaker)
\end{enumerate}

\begin{itemize}
\tightlist
\item
  If a tweet contains both criticism and antagonization, label as
  antagonization.
\item
  If you cannot confidently determine author stance (e.g., ambiguous
  sarcasm without clear target), prefer none.
\end{itemize}

\begin{enumerate}
\def\labelenumi{\arabic{enumi})}
\setcounter{enumi}{4}
\tightlist
\item
  Stance Detection Procedure
\end{enumerate}

\begin{itemize}
\tightlist
\item
  Step 1: Read the entire tweet (author's text only). Ignore potential
  quoted content for stance unless the author explicitly endorses it.
\item
  Step 2: Determine if the author personally expresses antagonization or
  criticism.

  \begin{itemize}
  \tightlist
  \item
    Antagonization: Personal attacks, dehumanization, us-versus-them or
    enemy framing, delegitimization.
  \item
    Criticism: Negative evaluation focused on actions, policies, or
    ideas, without personal attacks or enemy framing.
  \item
    None: Mere quoting/reporting/attribution, neutral statements,
    questions without negative stance, ambiguous irony with no clear
    stance.
  \end{itemize}
\item
  Step 3: Attribution rule:

  \begin{itemize}
  \tightlist
  \item
    If antagonizing/critical content is merely quoted, reported, or
    attributed (e.g., ``X meint\ldots'', ``laut \ldots'') with no
    endorsement, label none.
  \item
    If there is explicit endorsement (e.g., ``Genau!'', ``This!'',
    ``+1'', ``Stimme zu'', ``Agree''), treat as the author's stance and
    proceed to antagonization/criticism as appropriate.
  \end{itemize}
\end{itemize}

\begin{enumerate}
\def\labelenumi{\arabic{enumi})}
\setcounter{enumi}{5}
\tightlist
\item
  Target Identification (if antagonization or criticism)
\end{enumerate}

\begin{itemize}
\tightlist
\item
  Identify who/what is directly antagonized or criticized by the author.
\item
  Valid targets include:

  \begin{itemize}
  \tightlist
  \item
    Individuals (e.g., ``Habeck'', ``Trump'', ``@user'')
  \item
    Groups (e.g., ``Migranten'', ``Klimaaktivisten'', ``Medien'')
  \item
    Institutions/Entities (e.g., ``Regierung'', ``EU'', ``AfD'',
    ``Linke'', ``WHO'')
  \item
    Ideas/Policies (e.g., ``Klimaschutzgesetz'', ``Maskenpflicht'',
    ``Neoliberalismus'') when the critique is idea-focused.
  \end{itemize}
\item
  Quote the target exactly as it appears in the tweet. Do not translate.
\item
  If multiple targets are attacked/criticized, separate with a vertical
  bar: target1\textbar target2.
\item
  If the target is expressed vaguely, quote the exact phrase (e.g.,
  ``diese Leute'', ``ihr'', ``sie'', ``Idioten''). Do not infer or
  invent a more specific target from context.
\item
  Do not include entities that appear only:

  \begin{itemize}
  \tightlist
  \item
    as objects of someone else's quoted/attributed stance, or
  \item
    as neutral references unrelated to the author's
    antagonization/criticism.
  \end{itemize}
\end{itemize}

\begin{enumerate}
\def\labelenumi{\arabic{enumi})}
\setcounter{enumi}{6}
\tightlist
\item
  Special Cases and Decision Rules
\end{enumerate}

\begin{itemize}
\tightlist
\item
  Retweets and quote tweets:

  \begin{itemize}
  \tightlist
  \item
    Pure retweet (RT) or quote without author commentary: none.
  \item
    Quote tweet with commentary that endorses the quoted stance or adds
    a new stance: annotate based on the commentary.
  \end{itemize}
\item
  Hashtags:

  \begin{itemize}
  \tightlist
  \item
    Treat evaluative or pejorative hashtags as author stance (e.g.,
    ``\#Lügenpresse'' → antagonization; ``\#Versagen'' about a policy →
    criticism).
  \item
    Neutral/descriptive hashtags (e.g., \#Energie, \#News) do not imply
    stance.
  \end{itemize}
\item
  Mentions:

  \begin{itemize}
  \tightlist
  \item
    A mention (@user) is a target only if the author's stance is
    directed at the mentioned account.
  \end{itemize}
\item
  Emojis and stylings:

  \begin{itemize}
  \tightlist
  \item
    Insulting or demeaning emojis (e.g., clown, pile-of-poo) directed at
    a target contribute to antagonization.
  \item
    All-caps, repeated punctuation, or elongated words intensify tone
    but do not determine label by themselves.
  \end{itemize}
\item
  Sarcasm/Irony:

  \begin{itemize}
  \tightlist
  \item
    If sarcasm clearly conveys antagonization or criticism (e.g.,
    ``Toller `Plan', Regierung\ldots'' with negative context), label
    accordingly.
  \item
    If ambiguous (cannot identify stance or target), label none.
  \end{itemize}
\item
  Rhetorical questions:

  \begin{itemize}
  \tightlist
  \item
    If the question clearly expresses antagonization or criticism
    (``Wann hört diese inkompetente Regierung endlich auf?''), label
    accordingly.
  \item
    If the question is neutral or ambiguous, label none.
  \end{itemize}
\item
  Comparisons and metaphors:

  \begin{itemize}
  \tightlist
  \item
    Dehumanizing or hostile metaphors about people/groups/entities →
    antagonization.
  \item
    Metaphors about policies/ideas (e.g., ``Diese Strategie ist ein
    Irrweg'') → criticism.
  \end{itemize}
\item
  Policy vs person:

  \begin{itemize}
  \tightlist
  \item
    Attacking a person's character or legitimacy → antagonization.
  \item
    Critiquing actions, decisions, or policies → criticism.
  \end{itemize}
\item
  Generalization over groups:

  \begin{itemize}
  \tightlist
  \item
    Negative generalizations (``Ausländer ruinieren unser Land'') →
    antagonization.
  \end{itemize}
\item
  Negation and hedges:

  \begin{itemize}
  \tightlist
  \item
    ``Kein Fan von X'' can be criticism if the context evaluates an
    action/idea; can be antagonization if paired with insults or enemy
    framing.
  \end{itemize}
\item
  Ambiguity/unresolvable context:

  \begin{itemize}
  \tightlist
  \item
    If target or stance cannot be identified with reasonable certainty,
    label as ``none''.
  \end{itemize}
\end{itemize}

\begin{enumerate}
\def\labelenumi{\arabic{enumi})}
\setcounter{enumi}{7}
\tightlist
\item
  Output Fields for Each Tweet
\end{enumerate}

\begin{itemize}
\tightlist
\item
  stance: antagonization \textbar{} criticism \textbar{} none
\item
  target: string or null

  \begin{itemize}
  \tightlist
  \item
    If stance is antagonization or criticism, target must not be null.
  \item
    If multiple targets, separate with ``\textbar''.
  \item
    Quote target exactly as in the tweet; do not translate.
  \end{itemize}
\end{itemize}

\begin{enumerate}
\def\labelenumi{\arabic{enumi})}
\setcounter{enumi}{8}
\tightlist
\item
  Quick Decision Tree
\end{enumerate}

\begin{itemize}
\tightlist
\item
  Q1: Does the author themselves express a negative stance?

  \begin{itemize}
  \tightlist
  \item
    No → none.
  \item
    Yes → go to Q2.
  \end{itemize}
\item
  Q2: Is there an enemy/hostile framing or personal
  disparagement/insult?

  \begin{itemize}
  \tightlist
  \item
    Yes → antagonization.
  \item
    No → criticism (focused on actions/policies/ideas).
  \end{itemize}
\item
  Q3: Identify target(s) exactly as written.
\end{itemize}

\begin{enumerate}
\def\labelenumi{\arabic{enumi})}
\setcounter{enumi}{9}
\tightlist
\item
  Examples
\end{enumerate}

\begin{itemize}
\tightlist
\item
  Antagonization

  \begin{itemize}
  \tightlist
  \item
    Text: ``Die Medien lügen ständig und manipulieren das Volk.
    \#Lügenpresse''

    \begin{itemize}
    \tightlist
    \item
      stance: antagonization
    \item
      target: Medien\textbar\#Lügenpresse
    \item
      justification: Hostile framing and pejorative hashtag portray the
      media as manipulative enemies.
    \end{itemize}
  \item
    Text: ``Diese Politiker sind korrupt und gehören weg.''

    \begin{itemize}
    \tightlist
    \item
      stance: antagonization
    \item
      target: Politiker
    \item
      justification: Personal delegitimization and hostile removal
      framing indicate antagonization.
    \end{itemize}
  \item
    Text: ``Ihr seid Parasiten, die unser Land aussaugen.''

    \begin{itemize}
    \tightlist
    \item
      stance: antagonization
    \item
      target: ihr
    \item
      justification: Dehumanizing metaphor directed at a group signals
      antagonization.
    \end{itemize}
  \item
    Text: ``@user ist ein Clown ''

    \begin{itemize}
    \tightlist
    \item
      stance: antagonization
    \item
      target: @user
    \item
      justification: Direct personal insult and demeaning emoji.
    \end{itemize}
  \end{itemize}
\item
  Criticism

  \begin{itemize}
  \tightlist
  \item
    Text: ``Die Regierung hat in der Klimapolitik Fehler gemacht und
    sollte nachbessern.''

    \begin{itemize}
    \tightlist
    \item
      stance: criticism
    \item
      target: Regierung
    \item
      justification: Negative evaluation focused on policy performance
      without personal attacks.
    \end{itemize}
  \item
    Text: ``Das neue Gesetz ist unausgereift.''

    \begin{itemize}
    \tightlist
    \item
      stance: criticism
    \item
      target: Gesetz
    \item
      justification: Critique of a policy/idea, not of a person or
      group.
    \end{itemize}
  \item
    Text: ``Die EU sollte ihre Energie-Strategie überarbeiten.''

    \begin{itemize}
    \tightlist
    \item
      stance: criticism
    \item
      target: EU
    \item
      justification: Suggests policy change without hostile or demeaning
      framing.
    \end{itemize}
  \end{itemize}
\item
  None

  \begin{itemize}
  \tightlist
  \item
    Text: ``Einige Kritiker behaupten, die Regierung sei korrupt.''

    \begin{itemize}
    \tightlist
    \item
      stance: none
    \item
      target: null
    \item
      justification: The author reports others' views without adopting
      them.
    \end{itemize}
  \item
    Text: ``Alice Weidel behauptet, die Medien lügen ständig\ldots''

    \begin{itemize}
    \tightlist
    \item
      stance: none
    \item
      target: null
    \item
      justification: Pure reporting with no author commentary or
      endorsement.
    \end{itemize}
  \item
    Text: ``Morgen ist die Abstimmung im Parlament.''

    \begin{itemize}
    \tightlist
    \item
      stance: none
    \item
      target: null
    \item
      justification: Neutral informational statement.
    \end{itemize}
  \end{itemize}
\item
  Mixed and Tie-Breaker

  \begin{itemize}
  \tightlist
  \item
    Text: ``Die Regierung ist korrupt und ihre Steuerpolitik schadet
    der Mittelschicht.''

    \begin{itemize}
    \tightlist
    \item
      stance: antagonization
    \item
      target: Regierung
    \item
      justification: Contains a personal disparagement
      (``korrupt''), which takes precedence over policy critique.
    \end{itemize}
  \item
    Text: ``Schlechtes Urteil, aber die Richter sind keine Feinde.''

    \begin{itemize}
    \tightlist
    \item
      stance: criticism
    \item
      target: Urteil\textbar Richter
    \item
      justification: Evaluates a decision negatively while explicitly
      rejecting enemy framing.
    \end{itemize}
  \end{itemize}
\item
  Endorsement vs.~Attribution

  \begin{itemize}
  \tightlist
  \item
    Text: ``\,'Die Regierung ist korrupt.' -- laut einem Experten.''

    \begin{itemize}
    \tightlist
    \item
      stance: none
    \item
      target: null
    \item
      justification: Attributed to someone else; the author does not
      adopt the stance.
    \end{itemize}
  \item
    Text: ``\,'Die Regierung ist korrupt.' Genau so ist es.''

    \begin{itemize}
    \tightlist
    \item
      stance: antagonization
    \item
      target: Regierung
    \item
      justification: Explicit endorsement of a delegitimizing claim.
    \end{itemize}
  \end{itemize}
\end{itemize}

\begin{enumerate}
\def\labelenumi{\arabic{enumi})}
\setcounter{enumi}{10}
\tightlist
\item
  Quality and Consistency Tips
\end{enumerate}

\begin{itemize}
\tightlist
\item
  Be conservative with antagonization: require clear evidence of hostile
  framing or personal disparagement.
\item
  Do not infer targets beyond the text. Quote the target as written.
\item
  When in doubt about stance or target, label none and note the
  uncertainty in a ``note'' field
\item
  Apply the tie-breaker consistently: any personal attack or
  hostile/``enemy'' framing → antagonization.
\end{itemize}


\end{document}